\documentclass[a4paper,11pt]{article}
\pdfoutput=1
\usepackage{jheppub}
\usepackage{amsmath}
\usepackage{hyperref}
\usepackage{soul}
\allowdisplaybreaks
\def\bea#1\eea{\begin{align}#1\end{align}}
\def \be  {\begin{equation}}
\def \ee  {\end{equation}}
\newcommand{\nnu}{\nonumber}

\newcommand{\bef}{\begin{figure}[t]\centering}
\newcommand{\eef}{\end{figure}}
\newcommand{\sla}[1]{{#1}\!\!\!\slash}

\newcommand\as{\alpha_s}
\newcommand{\f}{\frac}

\newcommand{\ba}{\begin{eqnarray}}
\newcommand{\ea}{\end{eqnarray}}

\title{Jet angularity measurements for single inclusive jet production}

\author[a,b,c]{Zhong-Bo Kang,}
\author[d]{Kyle Lee}
\author[e]{and Felix Ringer}

\affiliation[a]{Department of Physics and Astronomy, University of California, Los Angeles, CA 90095, USA}
\affiliation[b]{Mani L. Bhaumik Institute for Theoretical Physics, University of California, Los Angeles, CA 90095, USA}
\affiliation[c]{Theoretical Division, Los Alamos National Laboratory, Los Alamos, NM 87545, USA}
\affiliation[d]{C.N. Yang Institute for Theoretical Physics, Stony Brook University, Stony Brook, NY 11794, USA}
\affiliation[e]{Nuclear Science Division, Lawrence Berkeley National Laboratory, Berkeley, CA 94720, USA}
                                   
\emailAdd{zkang@physics.ucla.edu}
\emailAdd{kunsu.lee@stonybrook.edu}
\emailAdd{fmringer@lbl.gov}

\abstract{We study jet angularity measurements for single-inclusive jet production at the LHC. Jet angularities depend on a continuous parameter $a$ allowing for a smooth interpolation between different traditional jet shape observables. We establish a factorization theorem within Soft Collinear Effective Theory (SCET) where we consistently take into account in- and out-of-jet radiation by making use of semi-inclusive jet functions. For comparison, we elaborate on the differences to jet angularities measured on an exclusive jet sample. All the necessary ingredients for the resummation at next-to-leading logarithmic (NLL) accuracy are presented within the effective field theory framework. We expect semi-inclusive jet angularity measurements to be feasible at the LHC and we present theoretical predictions for the relevant kinematic range. In addition, we investigate the potential impact of jet angularities for quark-gluon discrimination.}

\begin{document}
\maketitle

\section{Introduction}
\label{sec:intro}
Highly energetic jets and their substructure play a central role at present day hadron colliders like the Large Hadron Collider (LHC) and the Relativistic Heavy Ion Collider (RHIC). In the past years, the field of jet substructure has received a growing attention from both the experimental and theoretical communities. Applications of techniques involving jet substructure include precision tests of QCD, distinguishing quark and gluon jets, tagging of boosted objects and the search for physics beyond the standard model. To address these various applications, a range of different observables have been constructed in the past to describe and utilize the radiation pattern inside jets. See~\cite{Larkoski:2017jix} for a review of recent advances in applying jet substructure techniques to LHC physics.

In this paper we study jet angularities $\tau_a$ measured on an inclusive jet sample. Angularities were first introduced as a global event shape for di-jet events in $e^+e^-$ collisions~\cite{Berger:2003iw}. The parameter $a$ is a continuous variable, where for example $a=0,1$ correspond to thrust~\cite{Farhi:1977sg} and jet broadening~\cite{Catani:1992jc}, respectively. In~\cite{Almeida:2008yp}, jet angularities were proposed as a jet shape where the measurement is only performed on the constituents of a reconstructed jet. Studying a continuous class of jet shape observables generally allows for interesting insights into both the perturbative and non-perturbative structure of the jet dynamics~\cite{Berger:2003iw,Berger:2003pk,Berger:2003gr,Lee:2006fn}. The treatment within Soft Collinear Effective Theory (SCET)~\cite{Bauer:2000ew,Bauer:2000yr,Bauer:2001ct,Bauer:2001yt,Bauer:2002nz} for exclusive~\cite{Sterman:1977wj} $n$-jet events in $e^+e^-$ collisions was developed in~\cite{Ellis:2010rwa}. See also~\cite{Hornig:2009vb,Chien:2015cka,Bell:2017wvi} for example. The extension to exclusive di-jet events in $pp$ collisions was performed in~\cite{Hornig:2016ahz}. In general, jet substructure measurements can be performed on different jet sample. These include for example exclusive and inclusive di-jet production~\cite{CMS-PAS-SMP-16-010,Sun:2015doa,Aaboud:2017qwh} as well as single-inclusive jet production~\cite{Aad:2011kq,Aaboud:2017tke,ATLAS:2012am}. Exclusive jet production at $pp$ colliders always involves a veto $p_T^{\text{cut}}$ on the out-of-jet radiation within a given rapidity interval, see for example~\cite{Hornig:2017pud}. Instead, for inclusive jet production all jets in given rapidity $\eta$ and transverse momentum $p_T$ bins are taken into account and no further constraints are imposed on the event. Both the experimental and theoretical challenges can differ significantly when the jet substructure observable is measured using different event samples and the different approaches have advantages depending on the context. In this work, we focus on jet angularity measurements performed on an inclusive jet sample. Inclusive jet substructure observables allow for a simple and direct comparison between experimental data and first principle analytical results within QCD. In addition, single-inclusive jet substructure observables can be measured with the highest statistics  and they allow for a direct extension to heavy-ion collisions~\cite{Aad:2012vca,Abelev:2013fn,Chatrchyan:2013kwa,Aad:2014wha,Khachatryan:2016jfl,Kang:2017frl}. In this paper, we study specifically the ratio
\bea\label{eq:F}
F(\tau_a;\eta,p_T,R) = \left.\frac{d\sigma^{pp\to (\text{jet}\, \tau_a) X}}{d\eta dp_T d\tau_a}  \middle/ \f{d\sigma^{pp\to \text{jet} X}}{d\eta dp_T}\right.
\eea
where the numerator and denominator are the differential jet cross section with and without  the additional measurement of the angularity $\tau_a$. For the denominator of Eq. (\ref{eq:F}), we follow the formalism developed in~\cite{Kang:2016mcy} where a factorization formalism of the inclusive jet cross section in terms of hard functions and semi-inclusive jet functions (siJFs) was developed. This approach allows for the all order resummation of single logarithms in the jet size parameter $\alpha_s^n\ln^n R$ to next-to-leading logarithmic (NLL) accuracy. See also~\cite{Dasgupta:2014yra,Dai:2016hzf,Liu:2017pbb}. In order to calculate the numerator, we introduce different jet functions that appear in the factorization of the cross section. We will refer to this type of jet functions as semi-inclusive angularity jet functions (siAJFs), to reflect the fact that the angularity of an inclusively identified jet is measured. In $pp$ collisions, the factorized form of the cross section in the numerator of~(\ref{eq:F}) is given by
\bea\label{eq:fac1}
 \f{d\sigma^{pp\to (\text{jet}\,\tau_a) X}}{d\eta dp_T d\tau_a} = \sum_{abc} f_a(x_a,\mu) \otimes f_b(x_b,\mu) \otimes H_{ab}^c(x_a,x_b,\eta,p_T/z,\mu)\otimes {\cal G}_c(z,p_T, R, \tau_a,\mu)\;.
\eea
Here, $f_{a,b}$ denote the parton distribution functions (PDFs) in the proton with the corresponding momentum fraction $x_{a,b}$. The symbols $\otimes$ denote appropriate integrals over the variables $x_{a,b}$ and $z$. The hard functions $H_{ab}^c$ describe the production of an energetic parton $c$ in the hard-scattering event similar to inclusive hadron production~\cite{Aversa:1988vb,Jager:2002xm}. The new ingredient here are the siAJFs denoted by ${\cal G}_c(z,p_T R,\tau_a,\mu)$, which we are going to define at the operator level in the Sec.~\ref{sec:definition} below. Analogous to the siJFs~\cite{Kang:2016mcy}, the variable $z$ is the longitudinal momentum fraction of the parton $c$ initiating the jet that ends up inside the reconstructed jet. 

In order to allow for a meaningful comparison to experimental data, we need to resum two classes of logarithms to all orders. First, the resummation of small-$R$ logarithms is achieved by solving a DGLAP type evolution equation similar to the siJFs. Second, we need to resum logarithms of the form $\as^n \ln^{2n}(\tau_a^{\frac{1}{2-a}}/R)$, in the region where $\tau_a^{\frac{1}{2-a}} \ll R$. To that extend, we further demonstrate that the siAJFs can be refactorized in terms of soft and collinear functions $S_i(\tau_a,p_T,R,\mu)$ and $C_i(\tau_a,p_T,\mu)$, respectively, that describe the in-jet dynamics. This second step of the factorization is carried out at the jet scale $p_T R$ and it requires us to introduce further matching coefficients ${\cal H}_{c\to i}$ which describe the transition of the energetic parton $c$ coming from the hard-scattering event to the parton $i$ that initiates the jet. We obtain the following schematic structure
\bea\label{eq:refactorize}
{\cal G}_c(z,p_T, R,\tau_a,\mu) = \sum_i \mathcal{H}_{c\to i}(z,p_T R,\mu) \; C_i\left(\tau_a,p_T,\mu\right) \otimes S_i\left(\tau_a,p_T,R,\mu\right)\;,
\eea
where $\otimes$ represents a convolution over $\tau_a$ to be defined below. Note that upon integration of the siAJFs over $\tau_a$, we recover the siJFs at fixed order. Both steps of the outlined factorization theorem hold in the limit where the observed jet is sufficiently collimated. Therefore, we work with parametrically small values of the jet size parameter $R\ll 1$ even though for most practical purposes this is also a good approximation for e.g. $R\sim0.7$~\cite{Jager:2004jh,Mukherjee:2012uz} and even above. For large values of $R$, power corrections of the form ${\cal O}(R^2)$ can be systematically taken into account, see for example~\cite{Bertolini:2017efs}. Note that the structure of the refactorized semi-inclusive angularity jet function in Eq.~(\ref{eq:refactorize}) is very similar to~\cite{Kang:2017glf,Kang:2017mda} where (central) subjets and the transverse momentum distribution of hadrons inside jets were considered. However, here we are working within SCET$_{\text{I}}$, whereas in~\cite{Kang:2017glf,Kang:2017mda} the refactorized expression gave rise to collinear and soft modes on the same mass hyperbola which corresponds to SCET$_{\text{II}}$~\cite{Bauer:2002aj,Manohar:2006nz}. We would like to point out an important difference concerning the factorized structure in Eqs.~(\ref{eq:F}) and~(\ref{eq:refactorize}) and factorization theorems for exclusive jet production. In~\cite{Becher:2015hka,Chien:2015cka}, the authors introduced both a global soft and a soft-collinear (or `coft') mode in order to consistently separate all relevant modes and perform the all order resummation. For the calculation considered in this work, the situation is conceptually different since the out-of-jet radiation is not constrained to be small but instead it is unconstrained and fully taken into account in the two-step factorization procedure outlined above. See also for further inclusive jet substructure observables~\cite{Kaufmann:2016nux,Kang:2016ehg,Elder:2017bkd,Neill:2016vbi}.

Note that we do not take into account grooming in this work. Therefore, the obtained angularity is sensitive to non-global logarithms (NGLs)~\cite{Dasgupta:2001sh,Banfi:2002hw} due to the presence of the soft function obtained after the refactorization of the siAJFs in Eq.~(\ref{eq:refactorize}). While the extension to angularities with grooming is a separate task that is beyond the scope of this work, we would like to stress that ungroomed jet substructure observables play an important role for example in the context of heavy-ion collisions where a reliable background subtraction is necessary~\cite{Abelev:2013fn,Khachatryan:2016jfl,Adamczyk:2017yhe,Apolinario:2017qay}.

The remainder of this paper is organized as follows. In Sec.~\ref{sec:definition}, we present the factorized form of the cross section and we provide operator definitions for the siAJFs ${\cal G}_c(z,p_T, R,\tau_a,\mu)$. We calculate all relevant functions to next-to-leading order (NLO) and derive their renormalization group (RG) equations. By solving the obtained RG equations, we resum the relevant large logarithms to all orders in QCD. In addition, we demonstrate how the factorization for inclusive jet production is obtained upon integration of the siAJFs over $\tau_a$. In Sec.~\ref{sec:pheno}, we provide first numerical results for jet angularities measured on an inclusive jet sample for LHC kinematics. We include a shape function to model non-perturbative effects. Numerical results are presented for the potential application of jet angularities to quark-gluon discrimination. We summarize our work in Sec.~\ref{sec:summary} and provide an outlook.

\section{The semi-inclusive angularity jet function \label{sec:definition}}

We start by reviewing jet algorithms and angularities at hadron colliders~\cite{Almeida:2008yp,Hornig:2016ahz}. We then discuss the refactorized form of the siAJFs and we provide operator definitions for the collinear and soft functions. We present the corresponding perturbative results and discuss their renormalization and RG evolution. Finally, we show that the fixed order results for the siAJFs can be integrated over $\tau_a$ to obtain the siJFs and we discuss how the joint resummation of logarithms $\ln R$ and $\ln(\tau_a^{\f{1}{2-a}}/R)$ is achieved.

\subsection{Jet algorithms and angularity measurements at hadron colliders}

Here we briefly summarize the definition of jet angularity measurements at hadron colliders. For a more detailed discussion see~\cite{Almeida:2008yp,Hornig:2016ahz}. At NLO in $e^+e^-$ collisions, two final state particles are clustered together into the same jet when they satisfy the constraints
\bea
\text{cone jet : }& \theta_{iJ}<R\,,\\
k_T\text{-type}\text{ jet : }& \theta_{ij}<R\,.
\eea
Here $R$ is the jet size parameter, $\theta_{ij}$ is the angle between the particles $i$ and $j$ and $\theta_{iJ}$ is the angle between the jet axis and the particle $i$ that belongs to the jet. At hadron colliders jets are measured with a certain transverse momentum $p_T$ and rapidity $\eta$. Using the approximation that the highly energetic jets are sufficiently collimated, the implementation of the jet algorithm essentially amounts to replacing the jet parameter $R$ with
\bea
R \to \mathcal{R}\equiv\f{R}{\cosh \eta} \,.
\eea
The jet shape observables that we are interested in here are jet angularities which were defined in \cite{Almeida:2008yp,Ellis:2010rwa} as a jet shape for $e^+e^-$ colliders
\bea
\tau_a^{e^+e^-} = \f{1}{2E_J}\sum_{i\in J}|\vec{p}_{T}^{\;iJ}|\exp(-|\eta_{iJ}|(1-a))\;,
\eea
where $\eta_{iJ}$ is the pseudo-rapidity of the particles $i$ inside the jet and $\vec{p}_{T}^{\;iJ}$ denotes the transverse momentum measured with respect to the (standared) jet axis. The sum $i$ runs over all particles inside the reconstructed jet and $E_J$ is the jet energy. As mentioned in the Introduction, the parameter $a$ smoothly interpolates between different classic jet shape observables. As it was pointed out in \cite{Almeida:2008yp,Hornig:2016ahz}, hadron colliders prefer observables that are invariant under boosts along the beam direction. Therefore, the jet angularity for hadron colliders is defined as
\bea
\label{eq:tau_pp}
\tau_a\equiv\tau_a^{pp} \equiv \frac{1}{p_T} \sum_{i\in J} p_{T}^{i} \left(\Delta {\cal R}_{iJ}\right)^{2-a}
=  \left(\f{2E_J}{p_T}\right)^{2-a}\tau_a^{e^+e^-} + \mathcal{O}(\tau_a^2) \,, 
\eea
where $\Delta {\cal R}_{iJ} = \sqrt{(\Delta \eta_{iJ})^2+(\Delta \phi_{iJ})^2}$ with $\Delta\eta_{iJ}$ and $\Delta\phi_{iJ}$ the rapidity and azimuthal angle difference between the particle $i$ and the jet $J$. Note that the definition of $\tau_a$ also has a close relation to jet mass, $m_J$,
\bea
\tau_0 = \f{m_J^2}{p_T^2} + {\cal O}(\tau_0^2) \,.
\eea

\subsection{Factorization theorem}\label{sec:factorization}

Following~\cite{Kang:2017glf}, the siAJFs can be defined at the operator level for quark and gluon jets as follows
\bea
{\cal G}_q(z,p_T, R,\tau_a,\mu) &= \f{z}{2N_c}{\rm Tr} \left[\f{\sla{\bar n}}{2}
\langle 0| \delta\left(\omega - \bar n\cdot {\mathcal P} \right)\delta(\tau_a - \hat \tau_a(J)) \chi_n(0)  |JX\rangle \langle JX|\bar \chi_n(0) |0\rangle \right]\,,
\\
{\cal G}_g(z,p_T, R,\tau_a,\mu) &= - \f{z\,\omega}{2(N_c^2-1)}
\langle 0|  \delta\left(\omega - \bar n\cdot {\mathcal P} \right)\delta(\tau_a - \hat \tau_a(J)) {\mathcal B}_{n\perp \mu}(0) 
 |JX\rangle \langle JX|{\mathcal B}_{n\perp}^\mu(0)  |0\rangle\,,
\eea
where $\chi_n$ and ${\mathcal B}_{n\perp}^\mu$ are the gauge invariant collinear quark and gluon fields within SCET, and ${\cal P}$ is the label momentum operator. Here we have two light-like vectors $n^\mu = (1, \hat n)$ and $\bar n^\mu =(1, -\hat n)$ where $\hat n$ is aligned with the standard jet axis, and they satisfy $n^2=\bar n^2 =0$ and $n\cdot \bar n = 2$ as usual. In addition, $N_c$ is the number of colors for quarks, and the operator $\hat \tau_a(J)$ signifies the angularity measurement of the final observed jet, with the measured value equal to~$\tau_a$. Moreover, $\omega$ and $\omega_J$ are the large light-cone momentum components of the initial quark or gluon and the jet, with the ratio $ z =\omega_J / \omega$. Note that summation over the unobserved particles $X$ in the final is implied.

\bef
\includegraphics[width=3.1in]{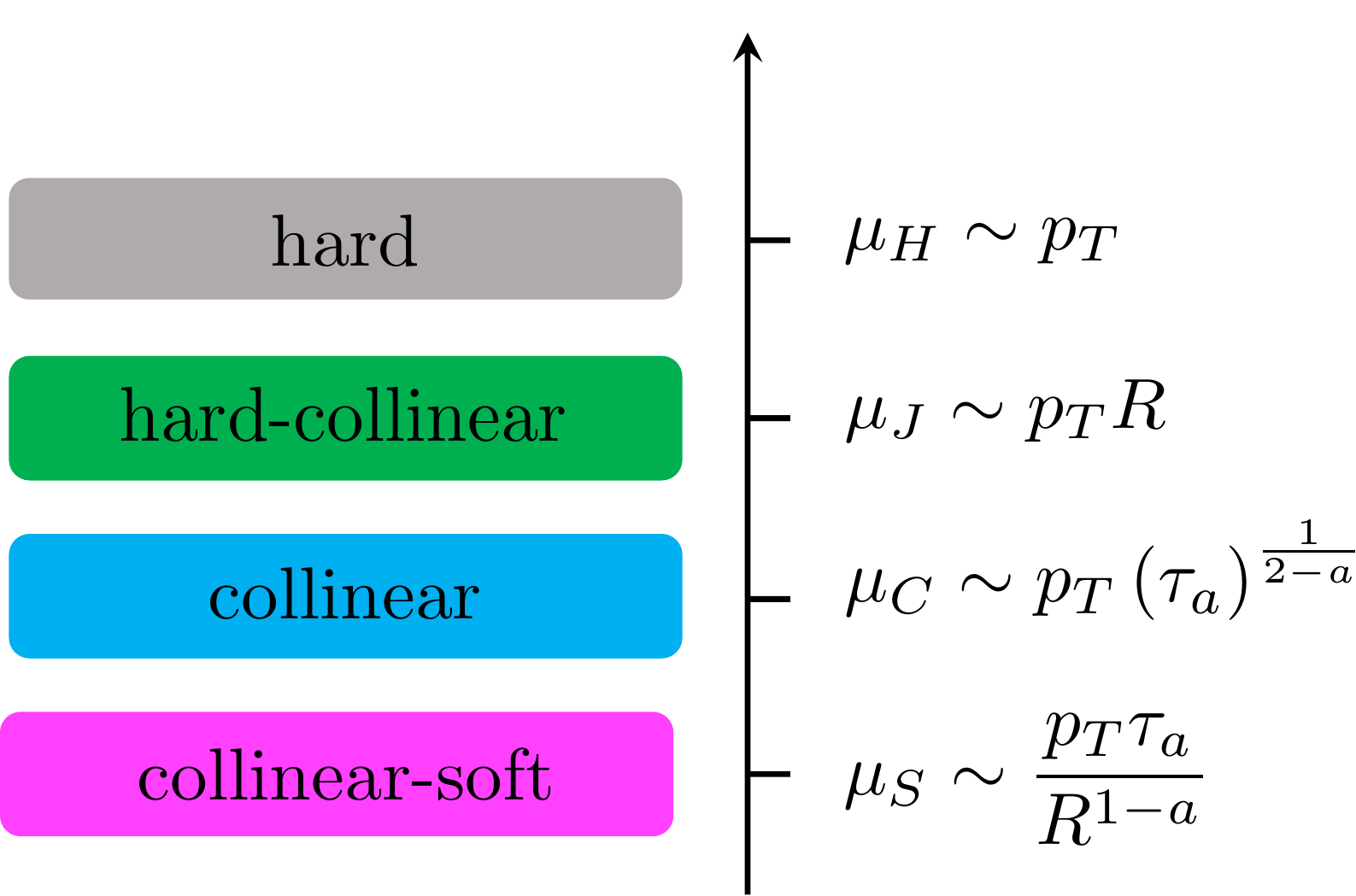} 
\caption{Characteristic momentum scales for all the relevant effective field theory modes for the factorization formalism in Eqs.~\eqref{eq:fac1} and \eqref{eq:refactorize2}.}
\label{fig:scales}
\eef

We are now going to discuss the factorization formalism for the jet angularity observable defined in Eq.~\eqref{eq:F} within SCET. The relevant effective field theory modes are summarized in Fig.~\ref{fig:scales}. The first step of the factorization in Eq.~\eqref{eq:fac1} is purely a separation of hard and collinear modes. The two relevant momentum scales are those associated with $H_{ab}^c$ and ${\cal G}_c$, respectively. The hard functions $H_{ab}^c$ have the characteristic momentum scale
\bea
\mu_H \sim p_T\,, 
\eea
whereas the characteristic momentum scale for the jet dynamics with a jet radius $R$, is given by
\bea
\mu_J \sim \omega_J\tan\left({\cal R}/2\right) \to p_T R\,. 
\eea
For the siAJFs ${\cal G}_c(z,p_T, R,\tau_a,\mu)$ there are in fact two relevant characteristic momentum scales, schematically $p_T \tau_a$ and $\mu_J\sim p_T R$. To be more precise, the relevant scale associated with $\tau_a$ should be given by $p_T\tau_a^{\f{1}{2-a}}$, see below. We will focus on the region where these two momentum scales are far separated, i.e. $\tau_a^{\f{1}{2-a}} \ll R$. In this case, an additional second step of the factorization as in Eq.~\eqref{eq:refactorize} is required in order to resum double logarithms of the form $\as^n \ln^{2n}(\tau_a^{\frac{1}{2-a}}/R)$. In this region, since $p_T\tau_a^{\f{1}{2-a}}$ is parametrically small, only collinear radiation within the jet with momentum scaling $p_c = (p_c^-, p_c^+, p_{c\perp})\sim p_T(1, \lambda^2, \lambda)$ with $\lambda\sim \tau_a^{\f{1}{2-a}}$, and the soft radiation of order $p_T\tau_a$ are relevant to leading power. 

An intuitive understanding of the collinear scaling $p_c$ can be obtained by realizing that the transverse momentum component $\lambda$ is roughly given by the typical angular separation $\theta_{iJ}$ of the collinear particles inside the jet with respect to the jet axis. From the definition of the jet angularity in Eq.~\eqref{eq:tau_pp}, one finds $\theta_{iJ}\sim \tau_a^{\f{1}{2-a}}$. Note that in the kinematic region $\tau_a^{\f{1}{2-a}}\ll R$ that we are considering, the collinear radiation is so collimated such that it is insensitive to the jet boundary~\cite{Hornig:2016ahz,Kang:2017glf}. Therefore, the collinear momentum scaling and the collinear function do not depend on the jet size parameter $R$. On the other hand, the precise momentum scaling for the soft radiation inside the jet is given by
\bea
p_s = \left(p_s^-, p_s^+, p_{s\perp}\right) \sim \frac{p_T\tau_a}{R^{2-a}}\left(1, R^2, R\right), 
\eea
which can be derived through the jet algorithm constraint $p_s^+/p_s^- \lesssim R^2$ and the definition of the jet angularity~\cite{Ellis:2010rwa}. Since soft radiation inside the jet only contributes to the observed jet angularities, we note that the soft degrees of freedom identified here are the same as the collinear-soft (c-soft) modes as in~\cite{Bauer:2011uc,Chien:2015cka}. Any harder emissions of the order $p_TR$ are only allowed outside the jet as they would otherwise break the hierarchy $\tau_a^{\f{1}{2-a}} \ll R$. They do not contribute to the angularity $\tau_a$ of the jet. We refer to modes taking into account such out-of-jet radiation as hard-collinear modes~\cite{Kang:2017mda,Kang:2017mda}, as labeled in Fig.~\ref{fig:scales}. In summary, in the limit $\tau_a^{\f{1}{2-a}} \ll R$ we obtain the following factorization structure for the siAJFs
\bea\label{eq:refactorize2}
{\cal G}_c(z,p_T, R,\tau_a,\mu) =& \sum_i \mathcal{H}_{c\to i}(z,p_T R,\mu) \nnu\\
&\hspace*{-2cm}\times\int d\tau_a^{C_i}d\tau_a^{S_i}\, \delta(\tau_a - \tau_a^{C_i}- \tau_a^{S_i})\, C_i\left(\tau_a^{C_i},p_T,\mu\right)
S_i\left(\tau_a^{S_i},p_T,R,\mu\right) \,,
\eea
where $C_i\left(\tau_a^{C_i},p_T,\mu\right)$ and $S_i\left(\tau_a^{S_i},p_T,R,\mu\right) $ denote collinear and soft functions that take into account collinear and soft radiation inside the jet. They both contribute to the angularity $\tau_a$ of the observed jet which is reflected by the convolution structure. For completeness, we provide the operator definitions here for both the collinear and soft functions. For the collinear functions, we have for quarks and gluons
\bea
C_{q}(\tau_a, p_T, \mu) =& \frac{1}{2N_c}
{\rm Tr} 
\bigg[\frac{\sla{\bar n}}{2}
\langle 0| \delta\left(\tau_a - \hat \tau_a^n \right)  \chi_n(0)  |JX\rangle
\langle JX|\bar \chi_n(0) |0\rangle \bigg]\,,
\\
C_{g}(\tau_a, p_T, \mu) =& - \frac{\omega}{2(N_c^2-1)}
\langle 0|  \delta\left(\tau_a - \hat \tau_a^{n} \right) {\mathcal B}_{n\perp \mu}(0) 
 |JX\rangle \langle JX|{\mathcal B}_{n\perp}^\mu(0)  |0\rangle\,.
\eea
Here the operator $\hat \tau_a^n$ is defined to count only the collinear radiation inside the jet. It is instructive to point out that these collinear functions are identical to the so-called measured jet functions in the context of exclusive jet production in~\cite{Ellis:2010rwa,Chien:2015cka,Hornig:2016ahz}. For the quark soft functions, we have 
\bea
S_q(\tau_a, p_T, R, \mu ) =& \frac{1}{N_c} \langle 0| {\bar Y}_n  \, \delta(\tau_a - \hat \tau_a^s)
Y_{\bar{n} }  | X\rangle \langle  X|{\bar Y}_{\bar n}  Y_n  |0\rangle,
\eea
where $Y_n$ is a soft Wilson line along the light-like direction $n^\mu$ of the jet, while $Y_{\bar n}$ is along the conjugated direction $\bar n^\mu$. Similar to the collinear function, the operator $\hat \tau_a^s$ is defined to count only the soft radiation inside the jet. The corresponding gluon soft functions is obtained by replacing the soft Wilson line by its counterpart in the adjoint color represenation and $N_c$ needs to be replaced with $N_c^2-1$ in the equation above. An important point worth mentioning is that the soft functions here only depend on two back-to-back directions, i.e. $n$ and $\bar n$. As pointed out in~\cite{Frye:2016aiz}, this can be understood in the sense that the collinear-soft modes obtained here are obtained from refactorizing jet functions, the siAJFs, which are color singlets. This relatively simple structure of the soft function is an important simplification compared to the more complex structure encountered for exclusive jet production.

As we are going to show below by explicit calculations, the natural momentum scales for the collinear and soft functions are given by
\bea
\mu_C\sim p_T (\tau_a)^{\f{1}{2-a}}\,,
\qquad
\mu_S\sim \f{p_T\, \tau_a}{R^{1-a}}\,.
\eea
On the other hand, $\mathcal{H}_{c\to i}(z,p_T R,\mu)$ are hard matching functions related to the harder radiation outside the jet, which have the natural momentum scale $\mu_J \sim p_T R$ and do not depend on $\tau_a$ as mentioned above.

\subsection{Hard matching functions \label{sec:hardmatching}}

The hard matching functions $\mathcal{H}_{c\to i}(z, p_T R, \mu)$ are obtained by matching onto the refactorized expression of the siAJFs in Eq.~(\ref{eq:refactorize2}). At NLO, they encode radiation that is of the order of the jet scale $\mathcal{O}(p_T R)$ which is only allowed outside of the jet in the kinematic region that we consider. They describe how an energetic parton $c$ coming from the hard-scattering event produces a jet initiated by parton $i$ with energy $\omega_J$ and radius $R$ carrying an energy fraction $z$ of the initial parton $c$. The same hard matching functions were obtained in the context of other jet substructure observables for inclusive jet production~\cite{Kang:2017mda,Kang:2017glf}. For both $k_T$-type and cone jets, the relevant expressions can be found in~\cite{Kang:2017mda}. The ${\cal O}(\alpha_s)$ expressions of the functions $\mathcal{H}_{c\to i}(z, p_T R, \mu)$ contain single and double logarithms of the form $L=\ln\left(\mu^2/p_T^2 R^2\right)$. These large logarithms vanish for the scale choice $\mu_J \sim p_T R$ which sets the initial scale for the evolution of the hard matching functions. After carrying out the renormalization, one finds the following RG equations
\bea
\mu \f{d}{d\mu}\mathcal{H}_{i \to j}(z,p_T R,\mu) = \sum_k \int_z^1 \f{dz'}{z'}\gamma_{ik}\left(\f{z}{z'},p_T R,\mu\right) \mathcal{H}_{k\to j}(z',p_T R,\mu) \,.
\eea
Note that the integro-differential structure of the evolution equations is similar to the standard DGLAP equations. However, here we have four coupled evolution equations $i\to j$. Also the anomalous dimensions differ from the usual DGLAP evolution kernels. We have
\bea\label{eq:gammaij}
\gamma_{ij}(z,p_T R,\mu) = \delta_{ij}\delta(1-z)\Gamma_i(p_T R,\mu) + \f{\alpha_s}{\pi}P_{ji}(z)\,,
\eea
where the second term are the usual Altarelli-Parisi splitting functions which resum single logarithms in the jet size parameter. The first (diagonal) term resums double logarithms. The coefficients $\Gamma_i(p_T R,\mu)$ contain a logarithmic term $\sim L=\ln\left(\mu^2/p_T^2 R^2\right)$ and are given by
\bea\label{eq:Gammaqg}
\Gamma_q(p_T R,\mu) = \f{\alpha_s}{\pi} C_F \left(-L-\f{3}{2}\right)\,,\\
\Gamma_g(p_T R,\mu) = \f{\alpha_s}{\pi} C_A \left(-L-\f{\beta_0}{2C_A}\right)\,.
\eea
To summarize, the RG equations encountered here resum single and double logarithms of the jet size parameter and the natural scale for the hard matching coefficients is given by the jet scale $\mu_J\sim p_T R$. Eventually, we are going to combine the hard matching functions at the jet scale with the collinear and soft functions in order to obtain the siAJFs ${\cal G}_i(z,p_T,R,\tau_a,\mu)$. In section~\ref{sec:res}, we demonstrate that the RG equations of the thus obtained siAJFs are again given by the usual DGLAP evolution equations associated with the resummation of single logarithms in the jet size parameter. This is the expected typical RG equation for jet substructure observables measured on an inclusive jet sample.

\subsection{Collinear functions \label{sec:collinear}}

The collinear functions $C_i\left(\tau_a,p_T,\mu\right)$ in the refactorized expression of the siAJFs in~(\ref{eq:refactorize2}) take into account collinear radiation inside the observed jet. The collinear functions are the same as encountered for exclusive jet production in~\cite{Ellis:2010rwa,Chien:2015cka,Hornig:2016ahz}. As we have emphasized in Sec.~\ref{sec:factorization}, to leading power, the collinear function is insensitive to the jet boundary and, hence, the value of $R$~\cite{Hornig:2016ahz,Kang:2017glf}. The jet algorithm constraint $\Theta_{\text{alg}}$ is only relevant for power corrections of the form $\mathcal{O}(\tau_a/R^2)$. The results for the collinear quark and gluon functions $i=q,g$ at NLO in $n=4-2\epsilon$ space-time dimensions are given by~\cite{Hornig:2009vb, Hornig:2016ahz} 
\bea\label{eq:pertcol}
C_i^{\text{bare}}(\tau_a,p_T,\mu) &= \delta(\tau_a) - \f{\alpha_s}{2\pi}\left[\left(\f{\mu^2}{p_T^2}\right)^{\epsilon}\left(\f{1}{\tau_a}\right)^{1+\f{2\epsilon}{2-a}}\left(\f{1}{\epsilon}\f{2 C_i}{1-a}+\f{\gamma_i}{1-a/2}\right)-\delta(\tau_a)f_i(a)\right] + \mathcal{O}\left(\f{\tau_a}{R^2}\right) \nonumber \\ 
&= \left\{1+\f{\alpha_s}{2\pi}\Big[f_i(a)+\f{\gamma_i}{\epsilon}+\f{C_i}{\epsilon^2}\f{2-a}{1-a}\Big]\right\}\delta(\tau_a) \nonumber \\
&-\f{\alpha_s}{2\pi}\Bigg\{ \left(\f{p_T}{\mu}\right)^{2-a}\left[\left(\f{\mu}{p_T}\right)^{2-a}\frac{1}{\tau_a}\right]_+\left(\f{1}{\epsilon}\f{2C_i}{1-a}+\f{\gamma_i}{1-a/2}\right)\nonumber \\
&-\f{4C_i}{(1-a)(2-a)}\left(\f{p_T}{\mu}\right)^{2-a}\left[\left(\f{\mu}{p_T}\right)^{2-a}\frac{1}{\tau_a}\ln\left(\tau_a\left(\f{p_T}{\mu}\right)^{2-a}\right)\right]_+\Bigg\} + \mathcal{O}\left(\f{\tau_a}{R^2}\right) \,.
\eea
Here we write the color factors as $C_i=C_{F,A}$ for quarks and gluons respectively. The functions $f_i(a)$ are given by
\bea
f_q(a) = \f{2C_F}{1-a/2}\Bigg[&\f{7-13a/2}{4}-\f{\pi^2}{12}\f{3-5a+9a^2/4}{1-a} \nonumber \\
&- \int_0^1 dx \f{1-x+x^2/2}{x}\ln\big[x^{1-a}+(1-x)^{1-a}\big]\Bigg]\;,
\\
f_g(a) = \f{1}{1-a/2}\Bigg[&C_A\Bigg((1-a)\Bigg(\f{67}{18}-\f{\pi^2}{3}\Bigg)-\f{\pi^2}{6}\f{(1-a/2)^2}{1-a} \nonumber \\
&- \int_0^1 dx \f{(1-x(1-x))^2}{x(1-x)}\ln\big[x^{1-a}+(1-x)^{1-a}\big]\Bigg)\nonumber \\
-T_R N_f&\Bigg(\f{20-23a}{18}-\int_0^1 dx (2x(1-x)-1)\ln\big[ x^{1-a}+(1-x)^{1-a}\big]\Bigg)\Bigg]\;,
\eea
and the constants $\gamma_i$ are
\bea
\gamma_q &= \f{3C_F}{2}\,, \hspace{0.5cm} \gamma_g = \f{\beta_0}{2} \,.
\eea
The results including power corrections can be found in \cite{Ellis:2010rwa}, which will be important in order to make the connection between the siAJFs and the siJFs as discussed in section~\ref{sec:integrate} below. Next, we consider the renormalization of the collinear functions. The bare and renormalized quantities are related as
\bea
C_i^{\text{bare}}(\tau_a,p_T) = \int d\tau_a' Z_{C_i}(\tau_a-\tau_a',p_T,\mu) C_i(\tau_a,\mu)\,.
\eea
The renormalization constants $Z_{C_i}$ are given by
\bea
Z_{C_i}(\tau_a,p_T,\mu) &= \left\{1+\f{\alpha_s}{2\pi}\Big[\f{\gamma_i}{\epsilon}+\f{C_i}{\epsilon^2}\f{2-a}{1-a}\Big]\right\}\delta(\tau_a) -\f{\alpha_s C_i}{(1-a)\pi}\f{1}{\epsilon} \left(\f{p_T}{\mu}\right)^{2-a}\left[\left(\f{\mu}{p_T}\right)^{2-a}\frac{1}{\tau_a}\right]_+\;,
\eea
and for the renormalized collinear functions we find
\bea
C_i(\tau_a,p_T,\mu) &= \left(1+\f{\alpha_s}{2\pi}f_i(a)\right)\delta(\tau_a)-\f{\alpha_s}{2\pi}\Bigg\{ \left(\f{p_T}{\mu}\right)^{2-a}\left[\left(\f{\mu}{p_T}\right)^{2-a}\frac{1}{\tau_a}\right]_+\left(\f{\gamma_i}{1-a/2}\right)\nonumber \\
&-\f{4C_i}{(1-a)(2-a)}\left(\f{p_T}{\mu}\right)^{2-a}\left[\left(\f{\mu}{p_T}\right)^{2-a}\frac{1}{\tau_a}\ln\left(\tau_a\left(\f{p_T}{\mu}\right)^{2-a}\right)\right]_+\Bigg\}\,.
\eea
From the renormalized expression, the natural scale of the collinear function can be obtained which is given by
\bea\label{eq:collinearscale}
\mu_C&\sim p_T (\tau_a)^{\f{1}{2-a}} \,,
\eea
where all large logarithms are eliminated at fixed order. The associated RG equation takes the following form
\bea\label{eq:collinearRG1}
\mu \f{d}{d\mu} C_i(\tau_a,p_T,\mu) = \int d\tau_a'\, \gamma_{C_i}(\tau_a - \tau_a',p_T,\mu)\, C_i(\tau_a',p_T,\mu) \,,
\eea
with the anomalous dimensions
\bea
\gamma_{C_i}(\tau_a,p_T,\mu) = \f{\alpha_s}{\pi}\left\{\left(C_i \f{2-a}{1-a}\ln\f{\mu^2}{p_T^2}+\gamma_i\right)\delta(\tau_a)-\f{2C_i}{1-a}\left(\f{1}{\tau_a}\right)_+\right\}\,.
\eea

\subsection{Soft functions \label{sec:soft}}

The soft functions $S_i\left(\tau_a,p_T,R,\mu\right)$ in Eq.~(\ref{eq:refactorize2}) take into account soft radiation within the identified inclusive jet. As mentioned above, the soft functions here correspond to the collinear-soft modes obtained in the context of exclusive jet production~\cite{Chien:2015cka}. The in-jet soft radiation directly contributes to the measured jet angularity $\tau_a$. Different than the collinear functions, they depend on the jet radius parameter $R$. To NLO, the soft functions for quarks and gluons~\cite{Ellis:2010rwa} are given by
\bea\label{eq:pertsoft}
S_i^{\text{bare}}(\tau_a,p_T,R,\mu) &= \delta(\tau_a)+\f{\alpha_s C_i}{\pi}\f{1}{\Gamma(1-\epsilon)}\left(\f{1}{1-a}\right)\f{1}{\epsilon}\f{1}{\tau_a^{1+2\epsilon}}\left(\f{\mu^2 e^{\gamma_E}R^{2(1-a)}}{p_T^2}\right)^\epsilon
\nonumber\\
&= \delta(\tau_a) + \f{\alpha_s C_i}{(1-a)\pi}\Bigg\{\f{\delta(\tau_a)}{2}\left(\f{\pi^2}{12}-\f{1}{\epsilon^2}\right)+\f{1}{\epsilon}\f{p_T}{\mu R^{1-a}}\left[\f{\mu R^{1-a}}{p_T \tau_a}\right]_+\nnu\\
&\hspace{1.3cm} - 2\f{p_T}{\mu R^{1-a}}\left[\f{\mu R^{1-a}}{p_T \tau_a}\ln\left(\f{p_T \tau_a}{\mu R^{1-a}}\right)\right]_+\Bigg\}
\,,
\eea
where we have omitted power corrections of the form $\mathcal{O}(R^2)$. Analogous to the collinear functions, the bare and renormalized soft functions are related in the following way
\bea
S_i^{\text{bare}}(\tau_a,p_T,R) = \int d\tau_a'\, Z_{S_i}(\tau_a-\tau_a',p_T,R,\mu)\, S_i(\tau_a,p_T,R,\mu) \,.
\eea
For the renormalization constants we find
\bea
Z_{S_i}(\tau_a,p_T,R,\mu) &= \delta(\tau_a) + \f{\alpha_s C_i}{(1-a)\pi}\Bigg(-\f{\delta(\tau_a)}{2\epsilon^2}+\f{1}{\epsilon}\f{p_T}{\mu R^{1-a}}\left[\f{\mu R^{1-a}}{p_T \tau_a}\right]_+\Bigg) \,,
\eea
and renormalized soft functions are given by
\bea\label{eq:softNLO}
S_i(\tau_a,p_T,R,\mu) &= \delta(\tau_a) + \f{\alpha_s C_i}{(1-a)\pi}\Bigg(\f{\pi^2}{24}\delta(\tau_a) - 2\f{p_T}{\mu R^{1-a}}\left[\f{\mu R^{1-a}}{p_T \tau_a}\ln\left(\f{p_T \tau_a}{\mu R^{1-a}}\right)\right]_+\Bigg) \,.
\eea
From this expression we can read off the natural scale of the soft function which is given by
\bea\label{eq:softscale}
\mu_S&\sim \f{p_T \tau_a}{R^{1-a}} \,.
\eea
The scale $\mu_S$ sets the starting scale for the evolution of the soft function. The renormalized soft functions follow the RG equation
\bea\label{eq:softRG1}
\mu \f{d}{d\mu} S_i(\tau_a,p_T,R,\mu) = \int d\tau_a' \gamma_{S_i}(\tau_a - \tau_a',p_T,R,\mu) S_i(\tau_a',p_T,R,\mu)
\eea
with the anomalous dimensions
\bea
\gamma_{S_i}(\tau_a,p_T,R, \mu) = \f{\alpha_s C_i}{\pi}\f{1}{1-a}\left[2\left(\f{1}{\tau_a}\right)_+-\ln\left(\f{\mu^2 R^{2(1-a)}}{p_T^2}\right)\delta(\tau_a) \right] \,.
\eea

\subsection{Integrating the semi-inclusive angularity jet function \label{sec:integrate}}

In this section, we demonstrate that the different functions of the refactorized siAJFs ${\cal G}_c(z,p_T, R,\tau_a,\mu)$ in Eq.~(\ref{eq:refactorize2}) can be integrated over $\tau_a$ in order to get back the siJFs $J_c(z,p_T R,\mu)$ which appear in the factorization theorem for inclusive jet cross section (or the $p_T$ spectrum)~\cite{Kang:2016mcy,Kaufmann:2015hma,Dai:2016hzf}. Note that the factorization theorem for the jet angularity differential distribution has a hard-collinear-soft structure in the kinematic regime discussed above in Eqs.~\eqref{eq:fac1} and \eqref{eq:refactorize2}. Upon integration over $\tau_a$, we need to get back to the inclusive jet cross section for which only a purely hard-collinear factorization is applicable. It is therefore interesting to study how this transition occurs when integrating out the $\tau_a$ dependence. In particular, the $\ln R$ dependence of the different functions is of interest and the obtained relation between the two cases may facilitate future higher order calculations for the inclusive jet spectrum. For exclusive jet production, a similar relation was obtained in~\cite{Chien:2015cka,Chay:2015ila} between the ``measured'' and ``unmeasured'' jet functions. The notion (un)measured jet function corresponds to jets where an additional measurement like the jet angularity is or is not performed. For exclusive jet production, it was found that soft and collinear pieces need to be combined correctly in order to obtain the ``unmeasured'' jet function from the ``measured'' one upon integration. For inclusive jet production the structure of the involved soft functions is much simpler as only in-jet soft radiation contributes. The out-of-jet radiation is unconstrained and integrated over both for the angularity differential case and inclusive jet production. As mentioned in the introduction, the $\tau_a$ differential cross section is calculated within SCET$_{\text{I}}$ like the inclusive jet cross section and a simple comparison of the singularity structure is thus possible. This is different than for example the inclusive jet substructure observables discussed in~\cite{Kang:2017mda,Kang:2017glf} where an additional rapidity regulator needs to be introduced (thus subtleties could arise~\cite{Collins:2003fm,GarciaEchevarria:2011rb}) which corresponds to SCET$_{\text{II}}$.

To simplify our discussion, we only consider the quark jet function and we choose $a=0$, to demonstrate
\bea
\label{eq:relation}
\int_{0}^\infty d\tau_0 \,{\cal G}_q(z,p_T R,\tau_0,\mu) = J_q(z,p_T R,\mu) \,.
\eea 
The refactorized form of the siAJFs ${\cal G}_c(z,p_T, R,\tau_0,\mu)$ in Eq.~(\ref{eq:refactorize2}) was derived in the limit $R\ll 1$ and $\tau_0\ll R^2$. Since also the siJFs are only known in the limit $R\ll 1$, we generally neglect power corrections of the form ${\cal O}(R^2)$. However, the second power counting used for our refactorization, $\tau_0\ll R^2$, requires that we include the first order power corrections of the form ${\cal O}(\tau_0/R^2)$ when we perform the integration over $\tau_0$. This is because the maximally allowed values for $\tau_0$ are given by~\cite{Ellis:2010rwa,Chien:2015cka}
\bea
\tau_0^{\text{max}}=  &\left\{
    \begin{array}{ll}
      \f{R^2}{4} \text{ for } k_T\text{-type}\,,\\ 
      R^2 \text{ for cone algorithms\,.} \\
    \end{array}
  \right.
\eea
Here we follow the procedure used in~\cite{Chien:2015cka} for exclusive jet production and we include the known one-loop power corrections when performing the integration. Alternatively, in~\cite{Chay:2015ila} the authors used a different power counting, $\tau_0\sim R^2$, when deriving the angularity measured cross section which can then be integrated up to the maximally allowed $\tau_0$. Note that only the collinear and soft functions discussed in sections~\ref{sec:collinear} and~\ref{sec:soft} above depend on $\tau_0$, whereas the hard matching coefficients ${\cal H}_{ij}$ of section~\ref{sec:hardmatching} are $\tau_0$ independent. The collinear function receives power corrections of the form ${\cal O}(\tau_0/R^2)$ whereas the soft function only has power corrections of order ${\cal O}(R^2)$. As an example, we consider the collinear quark function for $k_T$-type jets. Note that the same conclusions hold for cone jets. One has~\cite{Ellis:2010rwa}
\bea\label{eq:collinearpower}
C_q(\tau_0,p_T,R,\mu) = C_q^{\text{l.p.}}(\tau_0,p_T,\mu) + \Delta C_q^{\text{alg}}(\tau_0,R)\;,
\eea
where the first term is the leading power contribution as indicated by the superscript. For completeness, we present here the NLO result for $a=0$ which can be obtained from Eq.~(\ref{eq:pertcol}) in section~\ref{sec:collinear} above
\ba
C_q^{\text{l.p.}}(\tau_0,p_T,\mu) &= & \delta(\tau_0)+\f{\alpha_s C_F}{2\pi}\left\{\delta(\tau_0)\left(\f{3}{2\epsilon}+ \f{2}{\epsilon^2} + \f{7}{2}-\f{\pi^2}{2}\right) - \f{2}{\epsilon} \f{p_T^2}{\mu^2}\left[\f{\mu^2}{\tau_0 p_T^2}\right]_+ \right. \nnu\\
&& \left. -\f{3}{2}\f{p_T^2}{\mu^2}\left[\f{\mu^2}{\tau_0 p_T^2}\right]_++2\f{p_T^2}{\mu^2}\left[\f{\mu^2}{\tau_0 p_T^2}\ln\left(\f{\tau_0 p_T^2}{\mu^2}\right)\right]_+\right\} \,. 
\ea
For $k_T$-type algorithms, the power suppressed and algorithm dependent part for quarks at NLO is given by
\ba
\Delta C_q^{k_T}(\tau_0,R) &=& \f{\alpha_s C_F}{2\pi}\Bigg\{\f{\theta(\tau_0)\theta(\f{R^2}{4}-\tau_0)}{\tau_0} \left[3x_1 + 2\ln\left(\f{1-x_1}{x_1}\f{\tau_0}{R^2}\right)\right]\nnu\\
&&+\f{\theta(\tau_0 - \f{R^2}{4})}{\tau_0}\left(2 \ln \f{\tau_0}{R^2}+\f{3}{2}\right)\Bigg\} \,,
\ea
where 
\bea
x_1 = \f{1}{2}\left(1-\sqrt{1-\f{\tau_0}{\tau_0^{\text{max}}}}\right)\,.
\eea
The quark soft function at NLO for $a=0$ can be obtained from Eq.~(\ref{eq:softNLO}) and it is given by
\bea
S_i(\tau_0,p_T,R,\mu) =&\, \delta(\tau_0) + \f{\alpha_s C_i}{2\pi}\left\{\delta(\tau_0)\left( \f{\pi^2}{12}-\f{1}{\epsilon^2}\right)  +\f{2}{\epsilon} \f{p_T}{\mu R}\left[\f{\mu R}{\tau_0 p_T}\right]_+ \right. \nnu\\
& \left. - 4\f{p_T}{\mu R}\left[\f{\mu R}{p_T \tau_0}\ln\left(\f{p_T \tau_0}{\mu R}\right)\right]_+ \right\} \,.
\eea
By explicit calculation, one finds
\bea
\label{eq:part1}
\int_{\tau_0^{\text{max}}}^\infty d\tau_0 \,{\cal G}_q(z,p_T R,\tau_0,\mu) = 0 \,.
\eea
The results for the remaining integrals up to $\tau_0^{\text{max}}$ for the NLO collinear and soft quark functions for $k_T$-type jets are given by
\bea
\int_0^{\tau_0^\text{max}} d\tau_0\, C_q^{\text{l.p.}}(\tau_0,p_T,\mu) =&\, 1+ \f{\alpha_s C_F}{2\pi} \left\{\f{2}{\epsilon^2}+\f{3}{2\epsilon} -  \f{2}{\epsilon} \ln \left(\f{\tau_0^{\text{max}}p_T^2}{\mu^2}\right)+\ln^2\left(\f{\tau_0^{\text{max}} p_T^2}{\mu^2}\right)\right.\nnu\\
&\left.  -\f{3}{2}\ln\left(\f{\tau_0^{\text{max}} p_T^2}{\mu^2}\right) +\f{7}{2}-\f{\pi^2}{2}\right\}\,, \label{eq:colint} \\
\int_0^{\tau_0^\text{max}} d\tau_0\, \Delta C_q^{k_T}(\tau_0,R) =&\, \f{\alpha_s C_F}{2\pi} \left(3-\f{\pi^2}{3}-3\ln 2+4\ln^2 2\right) \,,\\
\int_0^{\tau_0^\text{max}} d\tau_0\, S_q(\tau_0,p_T,R,\mu)  =&\, 1+\f{\alpha_s C_F}{2\pi} \left\{- \f{1}{\epsilon^2} + \f{2}{\epsilon} \ln \left(\f{\tau_0^{\text{max}} p_T}{\mu R} \right)\f{\pi^2}{12}-2\ln^2\left(\f{\tau_0^{\text{max}} p_T}{\mu R}\right)\right\}\,.\nnu\\ \label{eq:softint}
\eea
When we sum over all contributions and use the maximally allowed value for $\tau_0$ for anti-$k_T$ jets, $\tau_0^{\text{max}}=R^2/4$, we obtain the in-jet contribution of the quark siJFs~\cite{Kang:2016mcy,Dai:2016hzf} which is equivalent to the ``unmeasured'' jet function for exclusive jet production~\cite{Ellis:2010rwa}. For completeness, we repeat the result here
\bea\label{eq:combine}
J_{q\to qg}(z,p_T R,\mu)=& \, \delta(1-z)\left[1+\f{\alpha_s}{2\pi}\left(\f{1}{\epsilon^2}+\f{1}{2\epsilon}-\f{1}{\epsilon}\ln\left(\frac{p_T^2 R^2}{\mu^2}\right)+\f{1}{2}\ln^2\left(\frac{p_T^2 R^2}{\mu^2}\right) \right. \right. \nnu \\
&\, \left.\left. -\f{3}{2}\ln\left(\frac{p_T^2 R^2}{\mu^2}\right)+\f{13}{2}-\f{3\pi^2}{4}\right) \right] \,.
\eea
Note that here we have only one type of logarithm left that can be eliminated at fixed order by choosing $\mu_J \sim p_T R$ which is the jet scale. As a last step, we can now combine this result with the expressions for the hard matching functions $\mathcal{H}_{q\to q} $ and $\mathcal{H}_{q\to g}$ which correspond to out-of-jet radiation diagrams at NLO. See for example Eqs.~(5.11) and~(5.12) of~\cite{Kang:2017mda}. We are then able to verify
\bea
\label{eq:part2}
\int_0^{\tau_0^\text{max}} d\tau_0\, {\cal G}_q(z,p_T, R,\tau_0,\mu) = J_q(z,p_T R,\mu) \,.
\eea
From Eqs.~\eqref{eq:part1} and \eqref{eq:part2}, we thus confirm the expected relation in Eq.~\eqref{eq:relation}. Note that after the integration over $\tau_0$, the collinear and soft functions contain $1/\epsilon^2$ poles in Eqs.~(\ref{eq:colint}) and~(\ref{eq:softint}) above. After combining them into a single function in Eq.~(\ref{eq:combine}), only one $1/\epsilon^2$ pole remains which is canceled by a corresponding term with opposite sign in the function ${\cal H}_{q\to q}$. We are thus left with only single poles and as well as single logarithms for the siJFs. By integrating over $\tau_0$ we have thus demonstrated explicitly how the hard-collinear-soft factorization theorem for the jet angularity distribution simplifies to the hard-collinear factorization of the inclusive jet cross section. Note that such a simplification does not occur for exclusive jet production where a hard-collinear-soft factorization is still required for the $\tau_0$ integrated result~\cite{Ellis:2010rwa,Chien:2015cka}.

\subsection{Resummation \label{sec:res}}

In this section, we perform the resummation of logarithms $\as^n \ln^{2n}(\tau_a^{\frac{1}{2-a}}/R)$ by solving the respective evolution equations of the collinear and soft functions. In addition, we demonstrate how the usual DGLAP equations are recovered for the evolution from the jet scale $\mu_J\sim p_T R$ to the hard scale $\mu\sim p_T$, which is associated with the resummation of single logarithms in the jet size parameter $\as^n\ln^n R$. First the collinear and soft functions are evolved to the jet scale $\mu_J$ starting from their respective natural scales. We then combine them with the hard matching functions of section~\ref{sec:hardmatching}. The evolution equations of the thus obtained siAJFs turn out to be the typical DGLAP equations where the anomalous dimensions are given by the Altarelli-Parisi splitting functions. All non-DGLAP terms of the anomalous dimensions cancel out between the different functions of the refactorized siAJFs. For all the relevant momentum scales, we refer to Fig.~\ref{fig:scales}. 

Here we choose to solve the evolution equations for the collinear and soft functions in Fourier transform space. See for example~\cite{Kang:2013wca}. We define the Fourier transform or position space expression of a generic function $F$ depending on $\tau_a$ as
\bea
F(x) = \int_0^\infty d\tau_a \, e^{-ix\tau_a} F(\tau_a) \,.
\eea
From Eqs.~(\ref{eq:pertcol}) and~(\ref{eq:pertsoft}), we obtain the following position space expressions for the bare collinear and soft functions at NLO
\bea
C_i^{\text{bare}}(x,p_T,\mu) &=\left\{1+\f{\alpha_s}{2\pi}\Big[f_i(a)+\f{\gamma_i}{\epsilon}+\f{C_i}{\epsilon^2}\f{2-a}{1-a}\Big]\right\} \nnu\\
&+\f{\alpha_s}{2\pi}\Bigg\{\ln\left(i\bar x\left(\f{\mu}{p_T}\right)^{2-a}\right)\left(\f{1}{\epsilon}\f{2C_i}{1-a}+\f{\gamma_i}{1-a/2}\right) \nnu \\
&+\f{2C_i}{(1-a)(2-a)}\left(\ln^2\left(i\bar x\left(\f{\mu}{p_T}\right)^{2-a}\right)+\f{\pi^2}{6}\right)\Bigg\}\;,
\\
S_i^{\text{bare}}(x,p_T,R,\mu) &= \left\{1 + \f{\alpha_s C_i}{2(1-a)\pi}\left(\f{\pi^2}{12}-\f{1}{\epsilon^2}\right)\right\}\nnu\\
&+\f{\alpha_s C_i}{(1-a)\pi}\Bigg\{-\f{1}{\epsilon}\ln \left( i\bar x\f{\mu R^{1-a}}{p_T}\right)
- \ln^2 \left( i\bar x\f{\mu R^{1-a}}{p_T}\right)-\f{\pi^2}{6}\Bigg\}\;,
\eea
where we introduced the shorthand notation $\bar x=xe^{\gamma_E}$. The convolution products as for example in Eqs.~(\ref{eq:collinearRG1}) and~(\ref{eq:softRG1}) turn into simple products in position space. We can thus write the RG equations for the collinear and soft functions as
\bea
\mu\frac{d}{d\mu}C_i(x,p_T,\mu) &= \gamma_{C_i}(x,p_T,\mu) \, C_i(x,p_T,\mu)\,,\\
\mu\frac{d}{d\mu}S_i(x,p_T,R,\mu) &= \gamma_{S_i}(x,p_T,R,\mu)\, S_i(x,p_T,R,\mu)\,.
\eea
The solution of these RG equations can be written as
\bea
\label{posevol}
C_i(x,p_T,\mu) &= \exp\left[\int^\mu_{\mu_{C}}\frac{d\mu'}{\mu'}\,\gamma_{C_i}(x,p_T,\mu')\right]C_i(x,p_T,\mu_C) \,,\\
S_i(x,p_T,R,\mu) &= \exp\left[\int^\mu_{\mu_{S}}\frac{d\mu'}{\mu'}\,\gamma_{S_i}(x,p_T,R,\mu')\right]S_i(x,p_T,R,\mu_S)\,,
\eea
where we evolved both functions to a common scale $\mu$ starting from their characteristic scales $\mu_{C,S}$, see Eqs.~(\ref{eq:collinearscale}) and~(\ref{eq:softscale}). The relevant anomalous dimensions are given by
\bea
\gamma_{C_i}(x,p_T,\mu)&= \f{\alpha_s}{\pi}\left[\gamma_i+\f{2C_i}{1-a}\ln\left(i\bar x\left(\f{\mu}{p_T}\right)^{2-a}\right)\right]\,,\\
\gamma_{S_i}(x,p_T,R,\mu)&=-\f{2\alpha_s C_i}{\pi}\f{1}{1-a}\left[\ln\left(i\bar x\f{\mu R^{1-a}}{p_T}\right) \right]\,.
\eea
Instead of evolving the collinear and soft functions separately to the hard scale $\mu\sim p_T$, we instead evolve only to the jet scale $\mu_J\sim p_T R$ where they are combined with the hard matching coefficients. We can thus write the distribution space expression for the evolved collinear and soft functions by taking the Fourier inverse transformation
\bea\label{eq:collinearsoftinverse}
&\int \f{dx}{2\pi} e^{ix\tau_a}C_i(x,p_T,\mu)S_i(x,p_T,R,\mu)= \int \f{dx}{2\pi} e^{ix\tau_a} \exp\left[\int^\mu_{\mu_{J}}\frac{d\mu'}{\mu'}(\gamma_{C_i}(x,p_T,\mu')+\gamma_{S_i}(x,p_T,R,\mu'))\right]\nnu\\
&\times\exp\left[\int^{\mu_J}_{\mu_C}\frac{d\mu'}{\mu'} \gamma_{C_i}(x,p_T,\mu')\right]\exp\left[\int^{\mu_J}_{\mu_S}\frac{d\mu'}{\mu'}\gamma_{S_i}(x,p_T,R,\mu')\right]C_i(x,p_T,\mu_C)S_i(x,p_T,R,\mu_S) \,.
\eea
Here we separated the evolution into two pieces. In the following, we demonstrate that the exponential in the first line that takes into account the evolution between the scales $\mu_J\to \mu$ cancels with a corresponding part of the evolved hard matching functions. The siAJFs then evolve according to the usual DGLAP evolution equations. Following~\cite{Kang:2017glf}, we can write hard matching functions $\mathcal{H}_{i\to j}(z,p_TR,\mu)$ as
\bea
\label{HEC}
\mathcal{H}_{i\to j}(z,p_T R,\mu) = \mathcal{E}_i (p_T R,\mu)\, \mathcal{C}_{i\to j}(z,p_T R,\mu)\,.
\eea
The functions $\mathcal{C}_{i\to j}(z,p_T R,\mu)$ follow evolution equations where the anomalous dimensions are given only by the Altarelli-Parisi splitting functions
\bea
\label{eq:Cevol}
\mu \f{d}{d\mu} \mathcal{C}_{i\to j} (z,p_T R,\mu) = \f{\alpha_s}{2\pi} \sum_k \int_z^1 \f{dz'}{z'} P_{ki}\left(\f{z}{z'}\right)\mathcal{C}_{k\to j}(z',p_T R,\mu)\,,
\eea
and the functions $\mathcal{E}_i(p_T R,\mu)$ satisfy multiplicative RG equations
\bea
\label{Eevol}
\mu \f{d}{d\mu} \ln \mathcal{E}_i (p_T R,\mu) = \Gamma_i(p_T R,\mu)\,.
\eea
Here the $\Gamma_i$ represent the purely diagonal pieces of the anomalous dimensions of the functions ${\cal H}_{i\to j}$ as given in Eq.~(\ref{eq:Gammaqg}). The fixed order results for both ${\cal C}_{i\to j}(z,p_TR,\mu)$ and ${\cal E}_{i}(p_T R,\mu)$ can be found in~\cite{Kang:2017glf}. The solution of the multiplicative RG equation for ${\cal E}_i(p_T,\mu)$ can be written as
\bea
\mathcal{E}_i(p_T R,\mu) = \mathcal{E}_i(p_T R,\mu_J) \exp\left(\int^\mu_{\mu_J} \f{d\mu'}{\mu'}\Gamma_i(p_T R,\mu')\right)\,.
\eea
Note that ${\cal E}_{i}(p_T R,\mu_J)=1$ when evaluated at the jet scale which sets the initial condition for the evolution. Moreover, we find that the remaining exponential factor from the evolution cancels with the corresponding part of the evolution of the collinear and soft functions between the scales $\mu_J\to \mu$ in Eq.~(\ref{eq:collinearsoftinverse}), i.e. we have
\bea
\exp\left(\int^\mu_{\mu_J} \f{d\mu'}{\mu'}\left[\Gamma_i(p_T R,\mu')+\gamma_{C_i}(x,p_T,\mu')+\gamma_{S_i}(x,p_T,R,\mu')\right]\right)=1\,.
\eea
To summarize, we can thus write the siAJFs ${\cal G}_c(z,p_T,R,\tau_a,\mu)$ in terms of the evolved functions as
\bea\label{eq:Gevol}
{\cal G}_c(z,p_T, R, \tau_a,\mu) =& \sum_i \mathcal{C}_{c\to i}(z,p_T R,\mu) \int \f{dx}{2\pi} e^{ix\tau_a} \exp\left[\int^{\mu_J}_{\mu_C}\frac{d\mu'}{\mu'} \gamma_{C_i}(x,p_T,\mu')\right]\nnu\\
&\times\exp\left[\int^{\mu_J}_{\mu_S}\frac{d\mu'}{\mu'} \gamma_{S_i}(x,p_T,R,\mu')\right]C_i(x,p_T,\mu_C)S_i(x,p_T,R,\mu_S) \,.
\eea
From Eq.~(\ref{eq:Cevol}) we find that the siAJFs follow the standard DGLAP evolution equations between the scales $\mu_J\to\mu$ which is associated with the resummation of single logarithms in the jet size parameter $R$.

\section{Phenomenology for $pp\to (\text{jet}\,\tau_a) X$ \label{sec:pheno}}

In this section, we present numerical result for the ratio $F(\tau_a;\eta,p_T,R)$ as defined in Eq.~(\ref{eq:F}) and repeated here for convenience
\bea
F(\tau_a;\eta,p_T,R) = \left.\frac{d\sigma^{pp\to (\text{jet}\, \tau_a) X}}{d\eta dp_T d\tau_a}  \middle/ \f{d\sigma^{pp\to \text{jet} X}}{d\eta dp_T}\right. \,.
\eea
The complete factorization theorem for the $\tau_a$ differential cross section can be found in Eq.~(\ref{eq:fac1}) above and the final result for the resummed siAJFs is given in Eq.~(\ref{eq:Gevol}). The single-inclusive jet cross section that appears in the denominator is obtained by replacing the siAJFs with the siJFs, i.e.
\bea
{\cal G}_c(z,p_T R, \tau_a,\mu)\to J_c(z,p_T R,\mu) \,.
\eea
Throughout this section, we work at NLL accuracy for the resummation of logarithms $\alpha_s^n\ln^n R$ and $\alpha_s^n\ln^{2n}(\tau_a^{\frac{1}{2-a}}/R)$. Note that in section~\ref{sec:res} above, we derived the resummed expressions in position space. For the numerical results presented in this section, we take the inverse transformation of the position space expression. As a cross check, we also performed the numerical calculations using an expression of the resummed result derived in momentum or distribution space and found full agreement. The $\ln R$ resummation is performed with the help of the numerical codes developed in~\cite{Vogt:2004ns,Anderle:2015lqa}.

\subsection{Non-perturbative shape functions and profile functions \label{sec:NPshape}}

\bef
\includegraphics[width=2.1in]{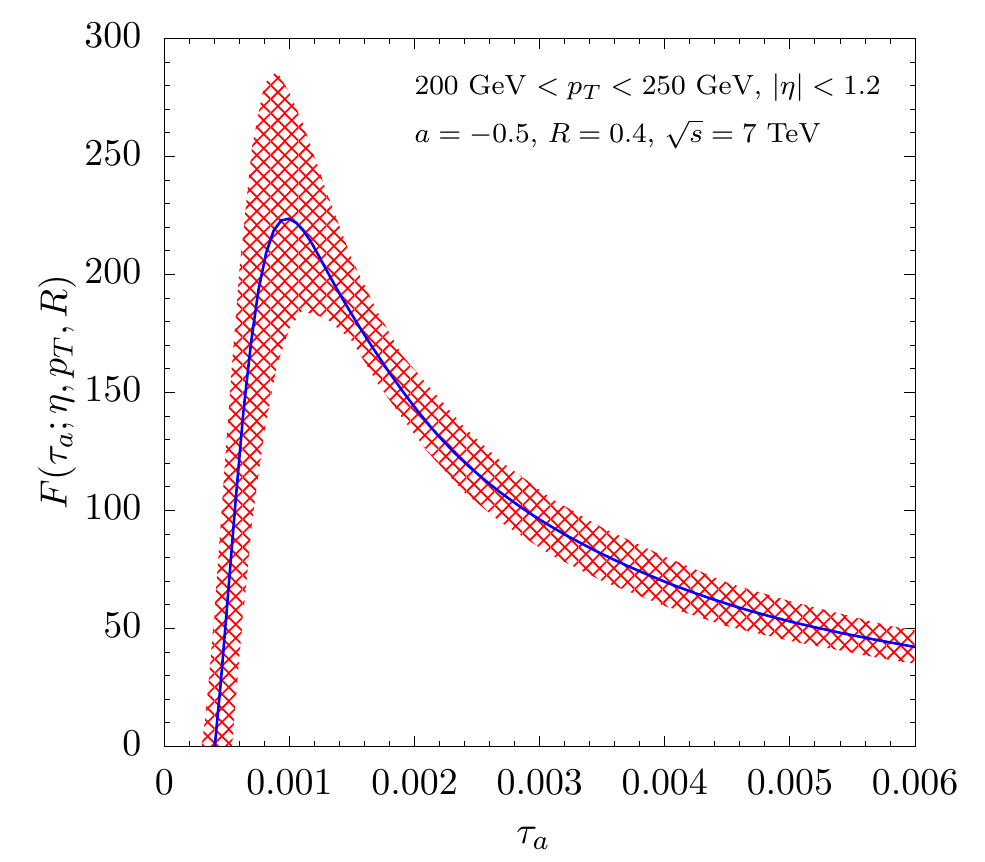} 
\hskip -0.25in
\includegraphics[width=2.1in]{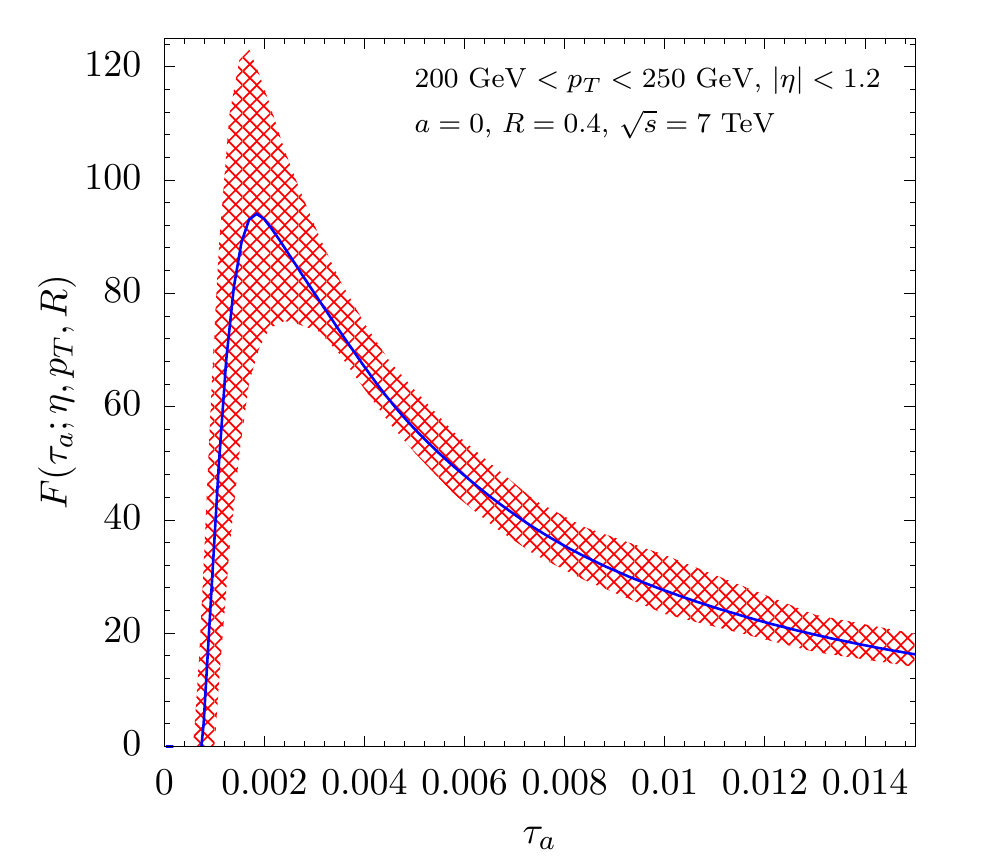} 
\hskip -0.25in
\includegraphics[width=2.1in]{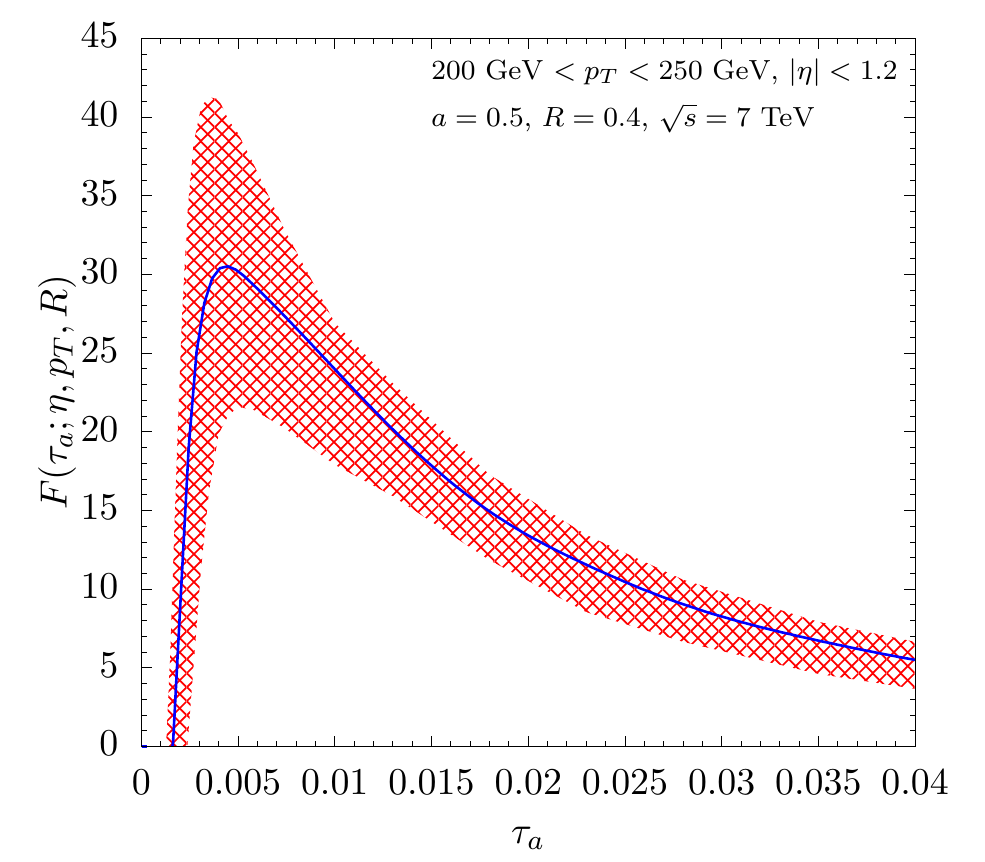} 
\caption{The jet angularity measured on inclusive jets in the $p_T$ range $200-250$~GeV at central rapidity $|\eta|<1.2$ at $\sqrt{s}=7$~TeV using the anti-$k_T$ clustering algorithm with $R=0.4$. As representative examples, we choose $a=-0.5,\, 0,\, +0.5$ from left to right. The scale uncertainty band is obtained as discussed in the text. \label{fig:jetangularity1}}
\eef

For small values of $\tau_a$, the soft scale $\mu_S\sim p_T \tau_a R^{a-1}$ could run into the non-perturbative regime. We parametrize this non-perturbative contribution with a shape function $S^{\text{NP}}(\tau_a)$. The new soft function is then given by a convolution of the purely perturbative result with $S_i^{\text{NP}}(\tau_a)$, i.e.
\bea
S_i(\tau_a,p_T,R,\mu_S) \to \int d\tau_a' \, S_i(\tau_a-\tau_a',p_T,R,\mu_S)\, S^{\text{NP}}_i(\tau_a')\,.
\eea
We adopt the following parametrization for the non-perturbative shape function~\cite{Kang:2013lga}
\bea\label{eq:shapefunction}
S_i^{\text{NP}}(\tau_a) = \f{\mathcal{N}(A,B,\Lambda)}{\Lambda}\left(\f{p_T \tau_a R^{a-1}}{\Lambda}\right)^{A-1}\exp\left(-\left(\f{p_T \tau_aR^{a-1}-B}{\Lambda}\right)^2\right) \,,
\eea
where $A, B,$ and $\Lambda$ are parameters. The ratio of the soft scale $\mu_S=p_T \tau_a R^{a-1}$ and $\Lambda$ determines where the non-perturbative effects start being important. The parameters $B$ and $A$ determine the location of the peak and the rate how fast the non-perturbative effects are turned off when the soft scale is in the perturbative regime, respectively. We make the following choices $\Lambda=0.4$, $A=2$, and $B=0.1$ for our numerical calculations as presented below. In the limit that the soft scale is far from the non-perturbative regime, or equivalently for large $\tau_a$, the cross section needs to approach the purely perturbative result. This is ensured by requiring that the non-perturbative shape function satisfy the following normalization condition
\bea
\int_0^\infty d\tau_a \, S_i^{\text{NP}}(\tau_a) = 1 \,,
\eea
from which the normalization factor $\mathcal{N}(A,B,\Lambda)$ in Eq.~(\ref{eq:shapefunction}) is determined. Note that we use the same shape functions for quarks and gluons. In order to ensure that $\alpha_s(\mu_S)$ does not run into the Landau pole for small values of $\tau_a$, we freeze the soft scale $\mu_S$ at some value above the Landau pole. This can be accomplished by making use of profile functions~\cite{Ligeti:2008ac}. We follow~\cite{Kang:2013lga}, by making the following choice to smoothly interpolate between regions I and II where the running of $\mu_S$ is turned off
\bea
f_{\text{profile}}(x)=&\left\{
    \begin{array}{ll}
      x_0[1+(x/x_0)^2/4] \hspace{1cm}x\leq 2x_0 \hspace{1cm}\text{region I}\,,\\ 
      x \hspace{3.7cm} x>2x_0  \hspace{1cm} \text{region II}\,.
    \end{array}
  \right.
\eea
\bef
\includegraphics[width=2in]{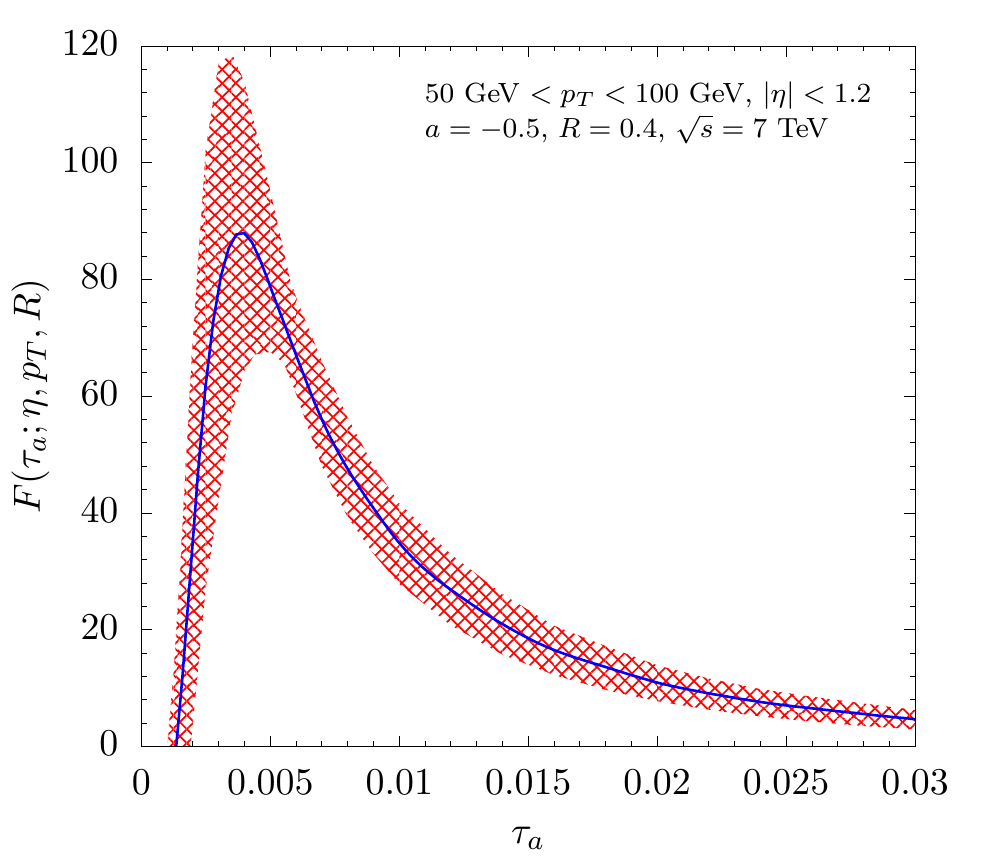} 
\hskip -0.15in
\includegraphics[width=2in]{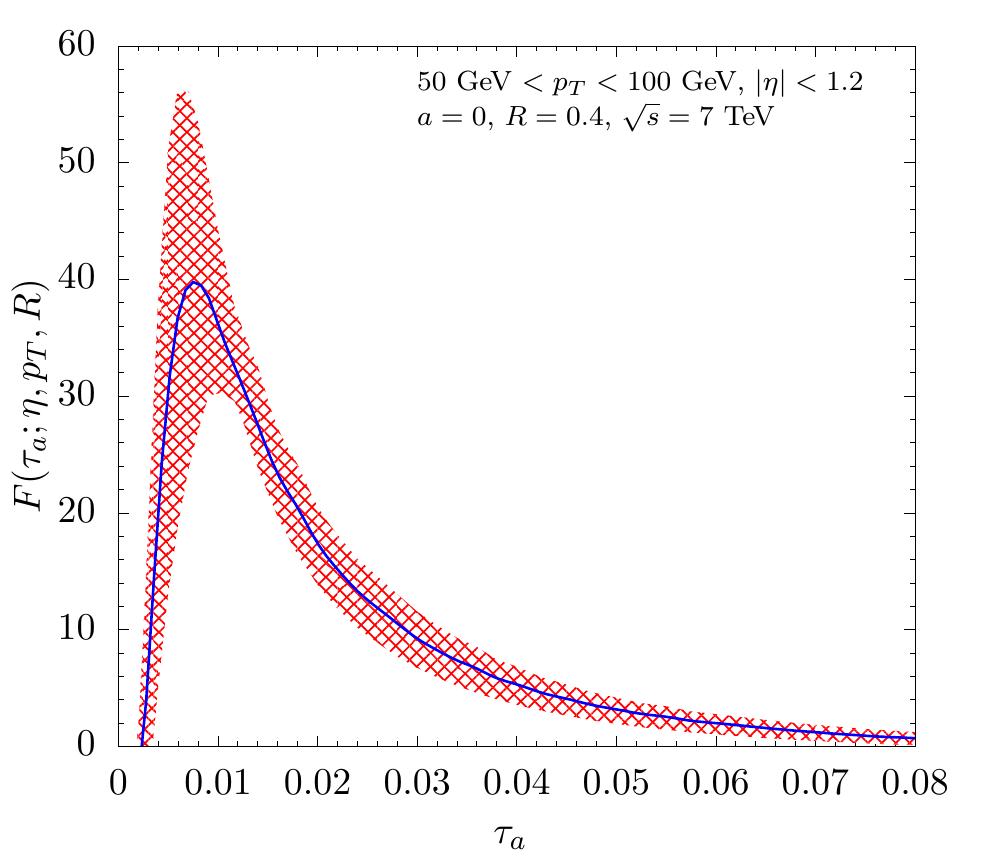} 
\hskip -0.15in
\includegraphics[width=2in]{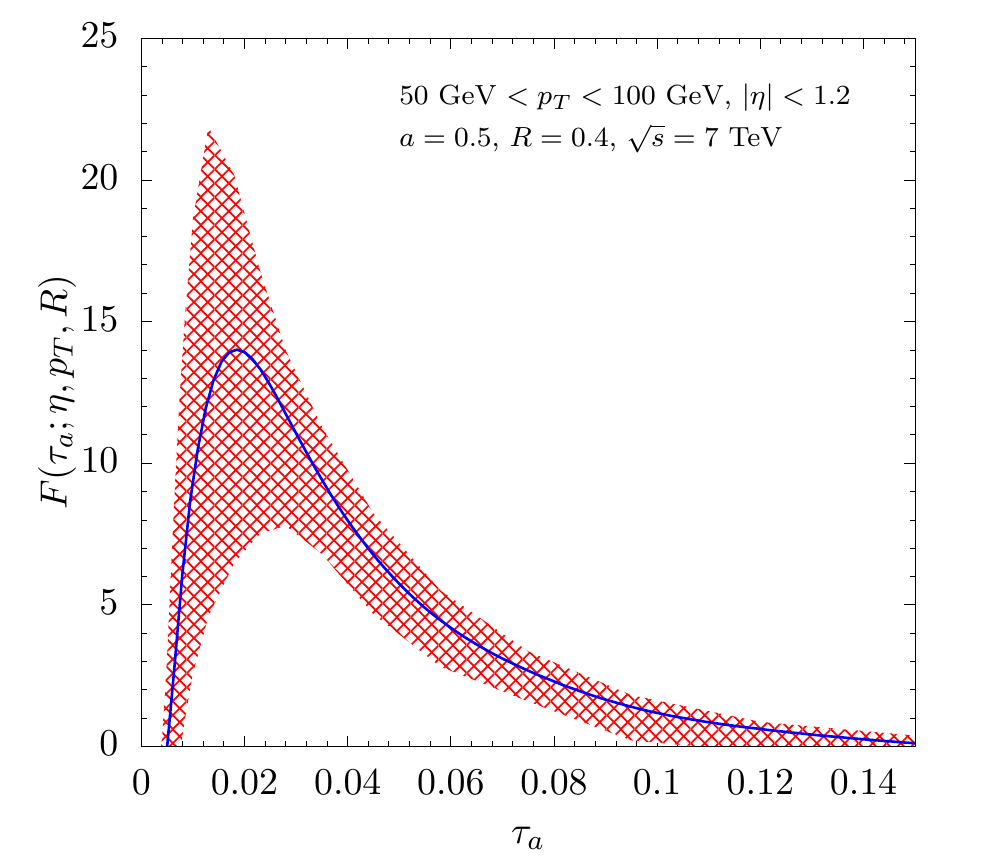} 
\caption{Same as Fig.~\ref{fig:jetangularity1} but for jets with a transverse momentum of $p_T=50-100$~GeV. \label{fig:jetangularity2}}
\eef
We then define our canonical scale choices for the soft and the collinear scale as
\bea\label{sc_rel}
\mu_S^{\text{can}} &= f_{\text{profile}}\left(\f{p_T \tau_a }{R^{1-a}}\right)\,,\\
\mu_C^{\text{can}} &= (\mu_S^{\text{can}})^{\f{1}{2-a}}(p_T R)^{\f{1-a}{2-a}}\,, 
\eea
where we make the choice
\bea
x_0 = 0.25\;\text{GeV}\,.
\eea
By making use of these profile functions, the value of the soft scale $\mu_S$ smoothly approaches the lower value $x_0$ and does not run into the Landau pole even in the limit $\tau_a\to 0$. Note that in Eq.~(\ref{sc_rel}) we wrote the canonical collinear scale choice $\mu_C^{\text{can}}$ in terms of the canonical choice for the soft scale $\mu_S^{\text{can}}$. In the next section, we discuss scale variations for which we always keep this relation between $\mu_C$ and $\mu_S$.

\subsection{Scale variations \label{sec:scalevariation}}

We vary the soft, jet and hard scales by factors of 2 around their canonical choices $\mu_S \sim p_T \tau_a R^{a-1}$, $\mu_J \sim p_T R$, and $\mu_H \sim p_T$ where we choose to keep the relation $1/2\leq (\mu_i/\mu_i^{\text{can}})/(\mu_j/\mu_j^{\text{can}})\leq 2$ between the different scales, where $i,j = S, J, H$. As mentioned above, the scale for the collinear function $\mu_C$ is varied together with soft scale $\mu_S$, see~Eq.~(\ref{sc_rel}). The variation of $\mu_S$ also must be turned off as $\mu_S$ approaches $x_0$. To freeze the variation of $\mu_S$, we define 
\bea
\mu_S = \left(1+ r \theta_\epsilon\left(p_T \tau_a R^{a-1}-2x_0\right)\right)\mu_S^{\text{can}} \,,
\eea
where the values $r = 0,-1/2,$ and $1$ correspond no variation, $1/2$, and $2$ times the canonical scale, respectively. The function $\theta_\epsilon$ is defined as~\cite{Hornig:2016ahz}
\bea
\theta_\epsilon\left(x-x'\right) = \f{1}{1+\exp[-(x-x')/\epsilon]} \,,
\eea
which approaches the standard theta function $\theta(x-x')$ in the limit $\epsilon \to 0$. For our numerical studies presented in the next section, we choose $\epsilon = 0.2~\text{GeV}$. This way, the variation of the soft scale $\mu_S \sim p_T \tau_a R^{a-1}$ is smoothly turned off when it is below the value of $2x_0$.

\subsection{Phenomenology at the LHC}

\bef
\includegraphics[width=2.8in]{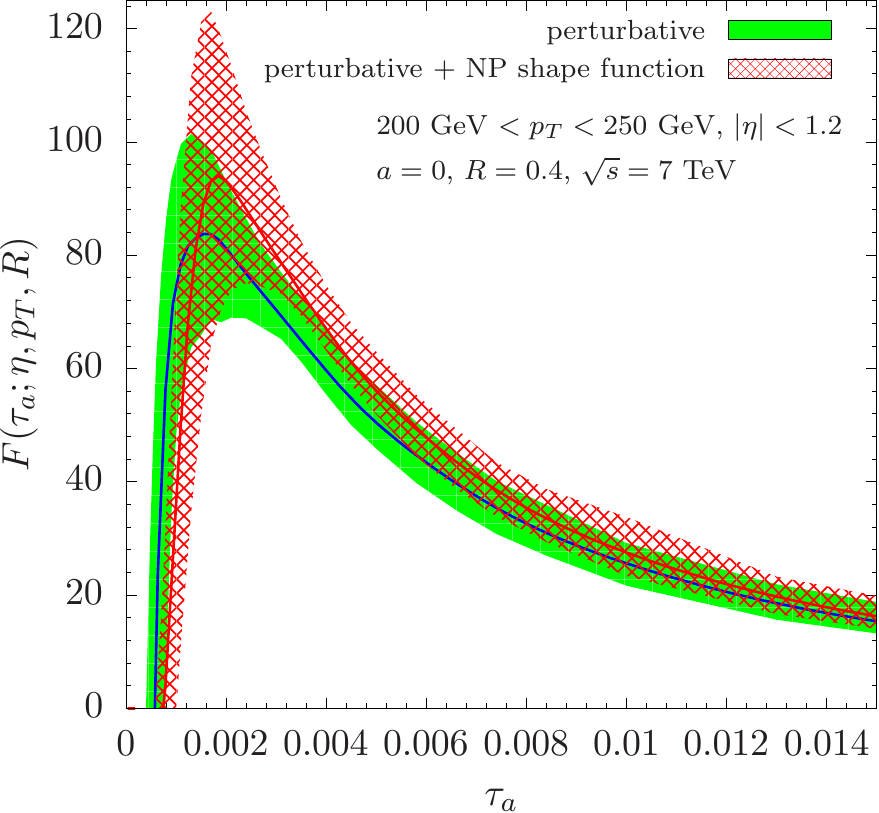}
\caption{The jet angularity distribution for $a=0$ with (red) and without (blue-green) the non-perturbative shape function. We use the same kinematical setup as in Fig.~\ref{fig:jetangularity1} as an example. \label{fig:wwoNP}}
\eef

We now present numerical results for the NLL resummed jet angularity distribution for inclusive jet production at the LHC $pp\to(\text{jet}\,\tau_a)X$. Throughout this section, we consider jets that are reclustered using the anti-$k_T$ algorithm~\cite{Cacciari:2008gp} and we use the CT14 PDF set of~\cite{Dulat:2015mca}. As an example, we consider a center of mass energy of $\sqrt{s}=7$~TeV and we require the observed jets to be at central rapidity $|\eta|<1.2$. In Fig.~\ref{fig:jetangularity1}, we present numerical results for inclusive jets in the transverse momentum range of $200<p_T<250$~GeV. In the three panels from left to right, we show the jet angularity distribution for $a = -0.5, 0,$ and $0.5$. Analogously, in Fig.~\ref{fig:jetangularity2} we show the results for jets with $50<p_T<100$~GeV. We include non-perturbative effects as outlined in section~\ref{sec:NPshape} above and the scale uncertainty bands are obtained following the discussion in section~\ref{sec:scalevariation}.

Jet angularity measurements capture different features of the radiation pattern inside a jet. The jet angularity measurement with a higher value of the parameter $a$ is more sensitive to collinear radiation. The increased sensitivity to collinear physics as $a\to 1$ causes the jet angularity to become sensitive to soft-recoil and the cross section cannot be factorized anymore when $a$ further increases to 2. For $a\geq 2$, the jet angularity cross section is infrared-collinear (IRC) unsafe. The growing sensitivity to collinear physics results in a larger scale uncertainty band. In other words, the cross section becomes less and less ``factorizable''. We also find that the height of the peak is reduced as $a$ increases. In addition, one generally finds that the distribution is peaked at smaller values of $\tau_a$ at higher jet transverse momenta. At smaller $p_T$, the jets are more dominated by gluons and they are broader. 
\bef
\includegraphics[width=2.1in]{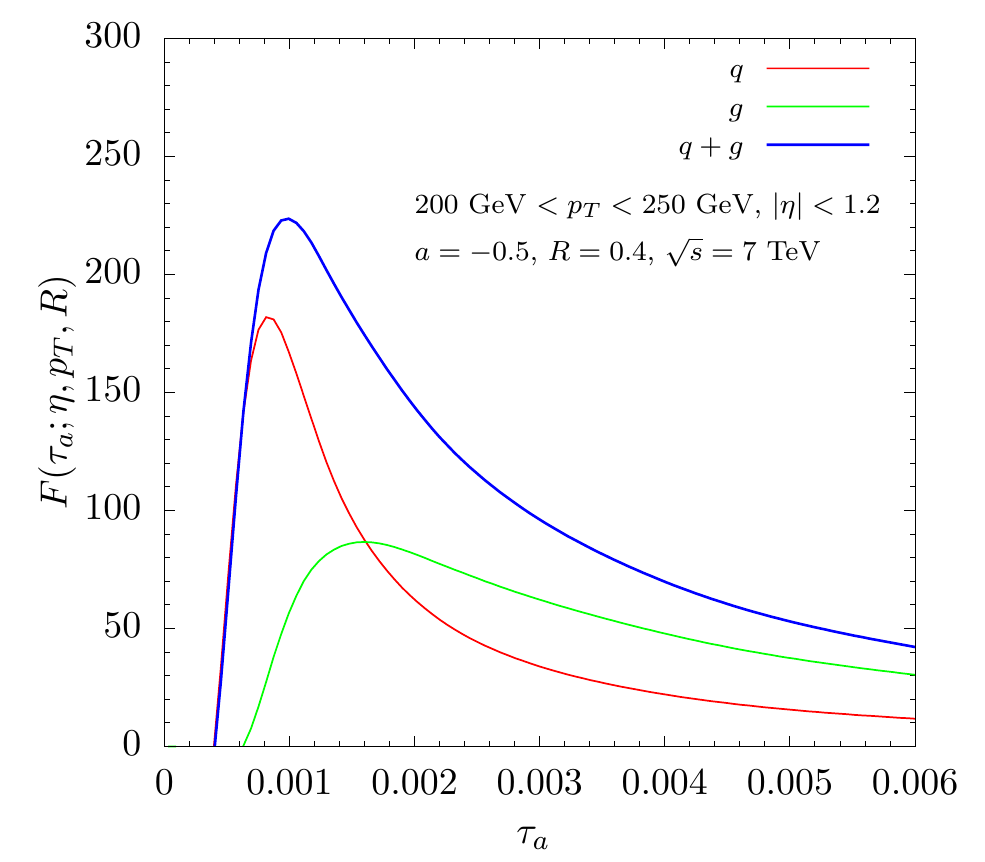} 
\hskip -0.25in
\includegraphics[width=2.1in]{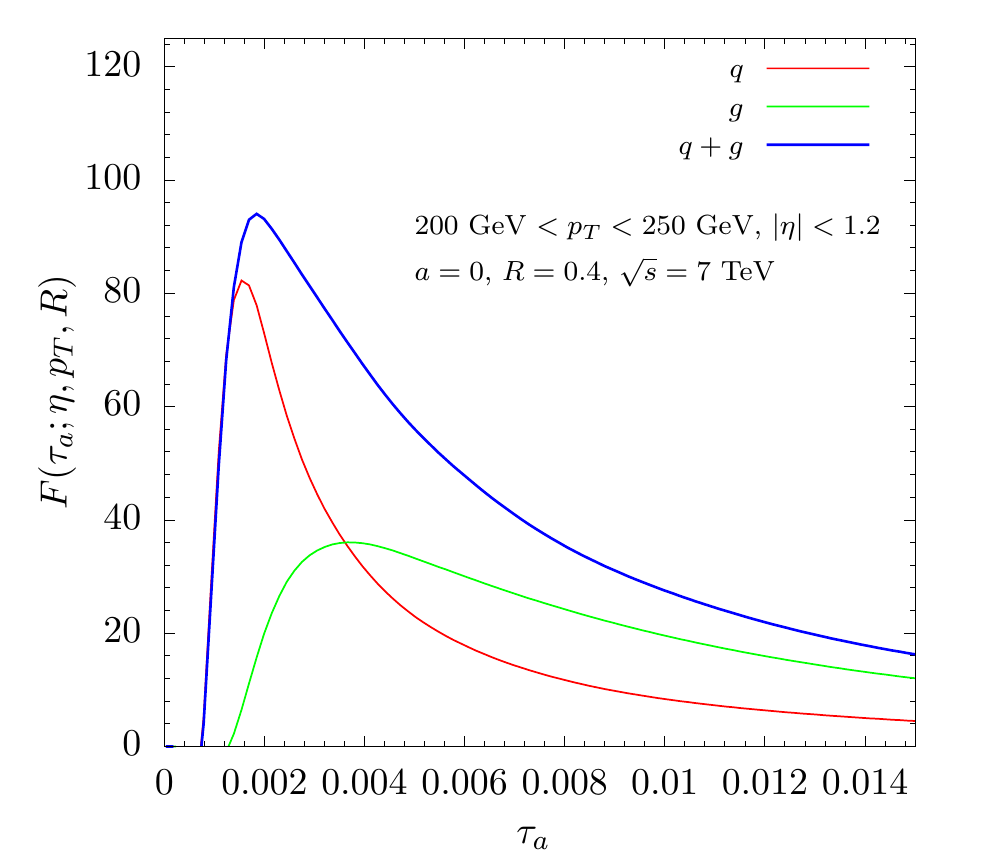} 
\hskip -0.25in
\includegraphics[width=2.1in]{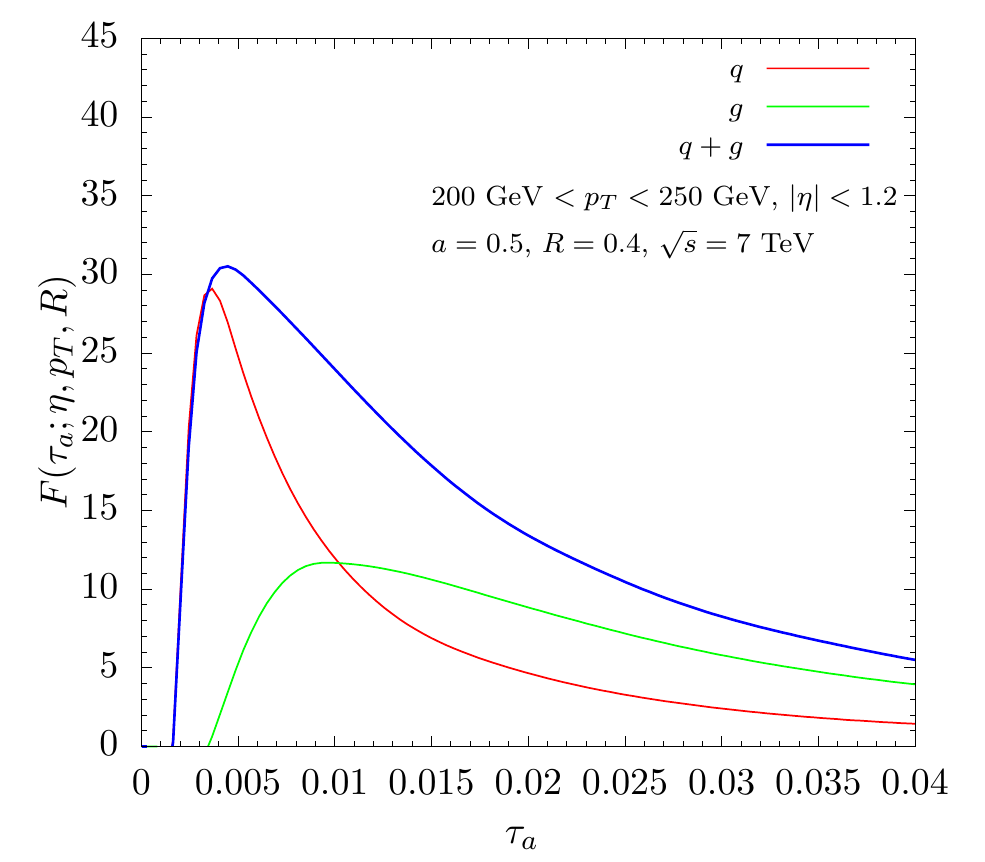} 
\caption{Jet angularities for the same kinematics as in Fig.~\ref{fig:jetangularity1} and $a=-0.5,\, 0,\, +0.5$ from left to right. We use the canonical scale choices. The total cross section (blue) is separated into quark (red) and gluon (green) contributions. \label{fig:quarkgluon1}}
\eef

Currently there is no data available for jet angularity distributions from the LHC that would allow us to determine the parameters of our model for the non-perturbative shape function from data. However, we expect that the corresponding measurements are feasile and that they can provide valuable information about non-perturbative dynamics and more generally about QCD factorization at present day hadron colliders. In order to gauge the relevance of non-perturbative effects, we show the result for the $a=0$ jet angularity distribution with (red) and without (blue-green) the non-perturbative shape function in Fig.~\ref{fig:wwoNP}. As an example, we use the same jet kinematics as in Fig.~\ref{fig:jetangularity1} and the details of the non-perturbative model were discussed in section~\ref{sec:NPshape} above. We observe a shift of the peak toward higher values of $\tau_a$. In addition, the height of the peak increases by roughly $\sim 10\%$ once the perturbative result is convolved with the non-perturbative shape function. The residual scale uncertainty band gets widened in particular in the peak region. In the tail region at large $\tau_a$, the two results converge as they should.

\subsection{Quark-gluon discrimination}

\bef
\includegraphics[width=5.8in]{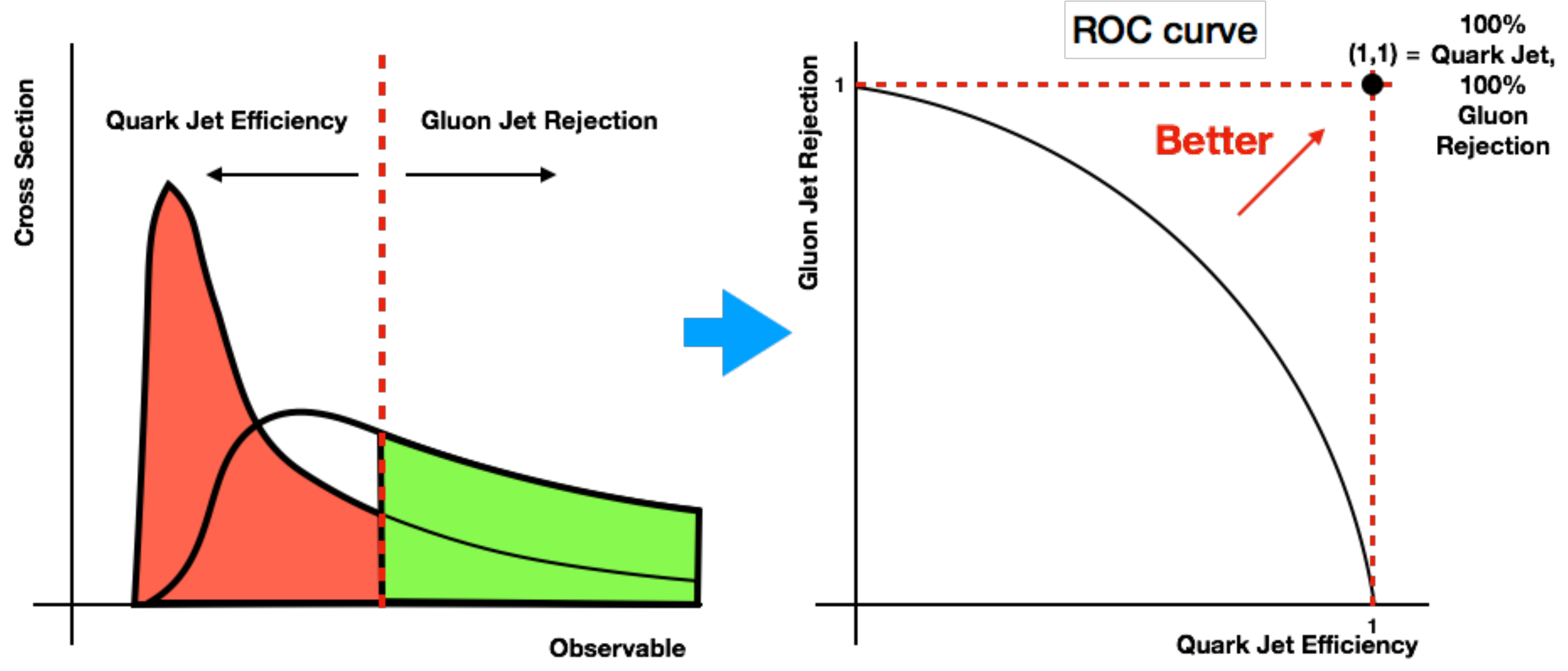}
\caption{As the sliding bar in the left figure moves through the values of some observable, here $\tau_a$, the amount of rejected background (gluon jets) and signal (quark jets) are recorded as points on the ROC curve in the right figure. The point $(1,1)$ in the upper right corner of the ROC curve plot corresponds to 100\% background rejection while keeping 100\% of the signal jets. \label{fig:disc}}
\eef

The discrimination of quark and gluon jets is an important goal of jet substructure techniques. One key motivation is that signatures of physics beyond the standard model at hadron colliders are often expected to be quark-heavy. See~\cite{Badger:2016bpw,Gras:2017jty} for an overview of quark-gluon tagging techniques. Modern classifiers include information from different IRC safe discriminant variables, hadron multiplicities or more recently also machine learning. See for example~\cite{Gallicchio:2012ez,Larkoski:2013eya,Larkoski:2014pca,Frye:2017yrw,Mo:2017gzp,Komiske:2016rsd,Metodiev:2017vrx,Butter:2017cot,FerreiradeLima:2016gcz,Cheng:2017rdo,Luo:2017ncs} and references therein. In this section, we study the potential applications of jet angularity measurements for quark-gluon discrimination from first principles analytical calculations in QCD. We start by separating the jet angularities into quark and gluon contributions. We show the cross section for the canonical scale choices in Fig.~\ref{fig:quarkgluon1} where the total cross section (blue) is separated into quark (red) and gluon (green) contributions. One notices that the gluon contribution shifted to larger values of $\tau_a$ is increased. As an example, we use the same kinematics as in Fig.~\ref{fig:jetangularity1}, i.e. $200<p_T<250$~GeV and $|\eta|<1.2$. We consider a Receiver Operating Characteristic (ROC) curve as illustrated in Fig.~\ref{fig:disc}. ROC curves show how well an observable discriminates between signal and background. Here we consider gluon jets as background and quark jets as signal. As the sliding bar separating quark jet efficiency and gluon jet rejection changes as a function of the observable, the resulting fractions are recorded on the ROC curve plot. As shown in the figure, the closer the ROC curve approaches the point $(1,1)$, the better the discrimination is between signal and background. An interesting aspect of using the jet angularities considered in this work is that we can study the quark-gluon discrimination efficiency as a function of the continuous variable $a$. See also~\cite{Gallicchio:2012ez}. In addition, since in- and out-of-jet radiation contributions are consistently taken into account in our framework, a direct comparison of data and analytical calculations from first principles in QCD is possible. The ROC curves for jet angularities are shown in Fig.~\ref{fig:ROC}. We show the result for four different values of $a=-0.5,\, 0,\, +0.5$ and $+0.8$. We observe that the quark-gluon discrimination is improved for $a\to 1$, where $a=1$ corresponds to the limit where the established factorization theorem breaks down. When $a\lesssim 1$, the jet angularity cross section is ``less under perturbative control'' and non-perturbative effects start to dominate. In fact, we find that the ROC curve for $a=0.8$ significantly depends on the non-perturbative model for the shape function as discussed in section~\ref{sec:NPshape}. We thus observe a tradeoff between having a better quark-gluon discriminant and the ability to perform (purely) perturbative calculations. For jet angularities, the transition between the two regions can be studied as a function of the continuous parameter $a$ and eventually an ideal intermediate value may be chosen.

\bef
\includegraphics[width=3.2in]{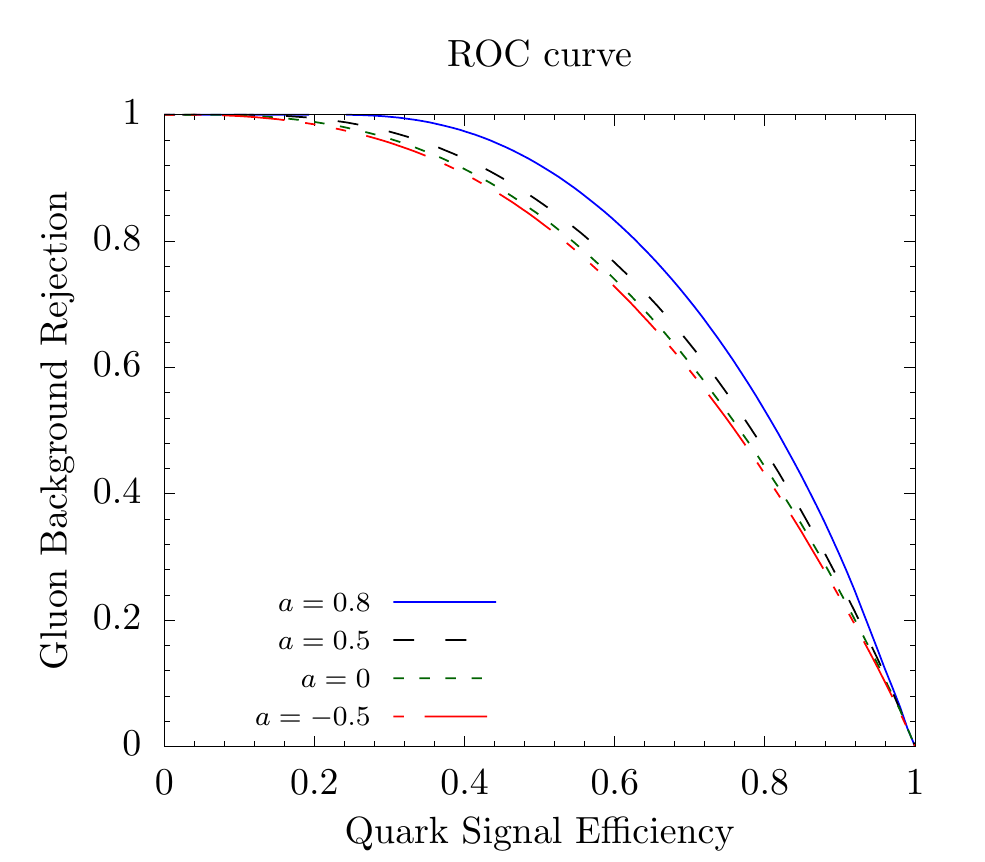} 
\caption{The ROC curve for jet angularities based on analytical first principles calculations in QCD for four different values of $a=-0.5,\, 0,\, +0.5,\, +0.8$. The jet angularities are measured on an inclusive jet sample with $200<p_T<250$~GeV and $|\eta|<1.2$ at $\sqrt{s}=7$~TeV as shown in Figs.~\ref{fig:jetangularity1} and~\ref{fig:disc} above. \label{fig:ROC}}
\eef

\section{Summary and outlook \label{sec:summary}}

In this work, we considered jet angularity measurements $\tau_a$ for inclusive jet production. We presented a corresponding factorization theorem, where the jet angularities are measured on an inclusive jet sample different than exclusive jet production considered earlier in the literature. All necessary functions were calculated to NLO which allowed us to determine the associated RG equations and anomalous dimensions. By solving the RG equations, we were able to jointly resum logarithms in the jet size parameter $R$ and the jet angularity $\tau_a$ to NLL accuracy. The obtained structure for the inclusive jet angularity measured cross section allowed for new insights also into the factorization theorem for inclusive jet production for which the relevant semi-inclusive jet function was obtained upon integration over $\tau_a$. We presented first numerical results for LHC kinematics for which we used profile functions and a shape function in order to systematically treat non-perturbative effects. We estimated the potential impact of jet angularities for quark-gluon discrimination by presenting ROC curves. We found that for larger values of $a\to 1$, the discrimination power between quark and gluon jets is improved while the sensitivity to non-perturbative effects is increased. In the future it will be worthwhile to study the impact of NGLs on the jet angularity distribution. By including NGLs, it will be possible to obtain the complete NLL resummed result. For example, in order to study the impact of NGLs, it will be interesting to compare jet angularities measured on both inclusive and exclusive jet sample. Another possible extension is to study groomed jet angularity distributions. The inclusive jet angularity distribution allows for a wide range of applications at the LHC including both proton-proton and heavy-ion collisions. We expect that the corresponding experimental measurements are feasible with the current and future data sets taken at the LHC. Finally, it will be interesting to explore applications of inclusive jet angularity measurements at RHIC and a future EIC~\cite{Accardi:2012qut,Aschenauer:2017jsk}.

\acknowledgments
We thank C.~Lee, X.~Liu, Y.~Makris, D.~Neill, N.~Sato and G.~Sterman for helpful discussions. This work is supported in part by the National Science Foundation under Grants No.~PHY-1316617, No.~PHY-1620628 and No.~PHY-1720486, the Department of Energy under Contract No.~DE-AC0205CH11231, and the LDRD Program of Lawrence Berkeley National Laboratory. 

\bibliographystyle{JHEP}
\bibliography{bibliography}

\end{document}